\documentclass[manuscript,preprint,rotate]{aastex}

\usepackage{latexsym,amsmath,amssymb}
\usepackage{natbib}
\usepackage{rotating}
\def \degmark{^\circ}
\def \nh {N${\rm _H}$}

\def \hcm {\hbox {\ifmmode $ atom cm$^{-2}\else atom cm$^{-2}$\fi}}
\def \arcmin {\hbox{$^\prime$}}

\def \ecut {\hbox {$E{\rm _{cut}}$}}
\def \chisq {$\chi ^{2}$}

\def\approxgt{\mathrel{\hbox{\rlap{\lower.55ex \hbox {$\sim$}}
        \kern-.3em \raise.4ex \hbox{$>$}}}}
\def\approxlt{\mathrel{\hbox{\rlap{\lower.55ex \hbox {$\sim$}}
        \kern-.3em \raise.4ex \hbox{$<$}}}}

\def \nineteen {XB\,1916$-$053}

\def \bigdip {X\,1624$-$490}

\def \exo {EXO\,0748$-$676}

\def \thirteen {XB\,1323$-$619}

\def \fouru {4U\thinspace1746$-$371}

\def \xiunit {\hbox{erg cm s$^{-1}$}}
\def \logxi {$\log(\xi)$}

\def  \comptt {{\tt compTT}}

\def \nh {$N{\rm _H}$}

\def \nhxabs {$N{\rm _H^{xabs}}$}

\shorttitle{INTEGRAL observations of four dipping LMXBs} 
\shortauthors{Balman \c{S}.}                                                

\begin{document}
\title{A study of the low-mass X-ray binary dip sources \nineteen, \thirteen, \bigdip\ and \fouru\ observed with
        INTEGRAL}

\author{\c{S}. Balman\altaffilmark{1,2}}

\altaffiltext{1} {Middle East Technical University, Physics Department, Inonu Bulvari,
       356531, Ankara, Turkiye; solen@astroa.physics.metu.edu.tr}
\altaffiltext{2} {Astrophysics Missions Division, Research and Scientific
       Support Department of ESA, ESTEC,
       Postbus 299, NL-2200 AG Noordwijk, The Netherlands
}

%\date{Received ; Accepted:}

\begin{abstract}

We detect dipping activity/modulations in the light curve of the four 
LMXBs in the 3--10 keV and 20--40 keV energy ranges.
The spectral parameters derived from the
fits to the INTEGRAL data are consistent with  hot coronal structures  in these
systems where we find a range of plasma temperatures  3.0--224.9
keV. The unabsorbed X-ray to soft
Gamma-ray flux between
4--200 keV are 5.9$\times 10^{-10}$ erg s$^{-1}$
cm$^{-2}$  for \nineteen, 3.3$\times 10^{-10}$ erg s$^{-1}$ cm$^{-2}$
for \thirteen, 21.6$\times 10^{-10}$ erg s$^{-1}$ cm$^{-2}$
for \bigdip\ and 11.0$\times 10^{-10}$ erg s$^{-1}$ cm$^{-2}$  for \fouru.
The optical depth to Compton scattering, $\tau$, varies in a range
4.4--0.002 consistent with electron densities 
$n{\rm _e}$ $<$ 1.4$\times 10^{15}$ cm$^{-3}$. 
%We recover a previously detected absorption line at 6.9$^{+1.8}_{-0.6}$~keV in 
%the JEM-X spectra of \nineteen\ and a new previously detected emission line
%at 6.8$^{+0.2}_{-0.2}$~keV in the JEM-X spectra of \bigdip.
In general, we find no significant difference in the dip and non-dip spectra 
in the ISGRI energy range (above 20 keV) for all the four sources.
We only detect absorption differences between dipping and non-dipping 
intervals for \nineteen\ and \bigdip\ in the JEM-X energy range.
Fits in the 4--200 keV range including an additional
photo-ionized absorber model for the two sources
show that \nineteen\  has the highest ionized absorber amoung the two.

\end{abstract} 


\section{INTRODUCTION}
\label{sect:introduction}

Some galactic low-mass X-ray binaries (LMXBs) show
periodic dips in their X-ray intensity.
The dipping behaviour repeats with the orbital period of the systems and is
caused by
periodic obscuration of the X-ray emitting region by a structure
located in the outer regions of a disk particularly belived to be
the impact region
of the accretion flow from the companion star
\citep{1916:white82apjl}. The depth, duration and spectral
properties of the dips vary according to the source and binary phase.
There are about 10 members (some candidates) of these systems.
The 1--10~keV spectra of most of the dip sources become
harder during dipping. This change is accepted to be inconsistent
with an  increase in the photo-electric absorption by cool
material since an excess of low-energy photons is also observed to be present
in some of the systems
\citep[e.g., \exo:][]{0748:cottam01aa}.

Several approaches is used to model this spectral
evolution (persistent--non-dip to dip spectra). The first approach is the "absorbed plus unabsorbed"
\citep[e.g., ][]{0748:parmar86apj, 1254:courvoisier86apj, smale:apj92} where the
persistent (non-dipping) spectral shape is used to model spectra
from dipping intervals.  The spectral evolution during dipping is
accounted for by large increases in the column density of the
absorbed component, and decreases of the normalization of the
unabsorbed component. The latter decrease is attributed to
electron scattering in the absorber. The changing normalization
was difficult to justify and thus, a more complete and updated version
of the model above was adopted as the "progressive
covering", or by name "complex continuum" approach \citep[e.g.,
][]{1916:church97apj, 1323:barnard01aa},
where the X-ray emission is
assumed to originate from a point-like blackbody (the neutron star), or
disk-blackbody component, together with an extended power-law
component. This approach uses a model
where the extended component is partially and progressively covered in time by an opaque absorber
to explain the spectral changes during dipping episodes. In such episodes,
the blackbody component is also absorbed.

More recently, the improved sensitivity and spectral resolution of
XMM-Newton, {\it Chandra} and $Suzaku$ have aided the detection of narrow absorption
features from highly ionized Fe (the Fe XXV and Fe XXVI lines)
and other metals from several X-ray binaries \citep{1915:lee02apj, 
1658:sidoli01aa, gx13:sidoli02aa, gx13:ueda04apj, 1624:parmar02aa, 1916:boirin04aa,
1916:juett06apj, gx339:miller04apj, 1743:miller06apj, 1630:kubotapsj07, Iaria:aa07, Iaria:apj06, 1254:diaztrigoaa09}.
A thorough analysis of XMM-Newton data of dipping LMXBs 
\citep{1323:boirin05aa, ionabs:diaz06aa} have recovered highly ionized Fe
absorption features from these systems and demonstrated that
modeling the changes in  both the observed X-ray absorption
lines and continuum seen during the dips and persistent emission can be
explained by an increase in the column
density and a decrease in the amount of ionization of a
highly photoionized absorbing plasma. 
%Since dipping sources are simply
%normal LMXBs viewed from close to the orbital plane, this indicates
%that photo-ionized absorbing regions may be found in several different type of LMXBs.

\begin{table}
\scriptsize{
\caption{Properties of the studied LMXBs. $L_{36}$
is the 0.6--10~keV luminosity in units of
10$^{36}$~erg~s$^{-1}$ for distances $d$ taken from
\citet{ionabs:diaz06aa}, except for \thirteen\ where the
0.5--10~keV luminosity is from \citet{1323:boirin05aa}.
\nhxabs\ is the persistent and dipping emission column  density in units of
$10^{22}$~atom~cm$^{-2}$ of the best-fit photo-ionized absorber
with an ionization parameter \logxi\ (in units of \xiunit)
measured by XMM-Newton \citep{1323:boirin05aa,
ionabs:diaz06aa}. Note that no such absorber was required with the
fits to \fouru\  \citep[]{ionabs:diaz06aa}.}

\begin{center}
\begin{tabular}[c]{lcccccr}
\hline \hline\noalign{\smallskip}
LMXB &  $P_{\rm dip}$ & $L_{36}$ & $d$ & \nhxabs & \logxi &$ P {\rm _{dip}}$\\
& (hr) & (erg s$^{-1}$) & (kpc)  &  Non-dip / Dip & Non-dip / Dip & Reference\\
\noalign{\smallskip\hrule\smallskip}
\nineteen & $0.833351411(25)$ & 4.4 & 9.3 & 4.2$^{+0.5}_{-0.5}$/28$^{+2}_{-2}$ & 3.05$^{+0.04}_{-0.04}$/2.55$^{+0.04}_{+0.04}$ & \citet{1916:chou01apj} \\
\thirteen & $2.94(2)$ & 5.2 & 10  & 3.8$^{+0.4}_{-0.4}$ / 14$^{+10}_{-7}$  & 3.9$^{+0.04}_{-0.04}$ / 3.0$^{+0.2}_{-0.2}$ & \citet{1323:balucinska99aa} \\
\fouru    & $5.16(1)$   & 10.1 & 10.7 & \dots & \dots & \citet{1746:balucinska04mn}  \\
\bigdip   & $20.8778(3)$  & 47.5 & 15   & 13$^{+2}_{-0.4}$ / 29$^{+4}_{-4}$ & 3.6$^{+0.2}_{-0.2}$ / 3.0$^{+0.2}_{-0.2}$ & \citet{1624:smale01apj} \\
\noalign{\smallskip\hrule\smallskip}
\end{tabular}
\end{center}
\label{tab:properties}
}
\end{table}

The dipping LMXB sources have been well studied with the RXTE and 
Beppo-Sax missions in an energy range out to $\sim$ 20 keV \citep{1746:parmar99aa, 1746:balucinska04mn, 1323:barnard01aa,
1323:balucinska99aa, 1916:narita03apj, 1916:church98aa, 1624:balucinska00aa, 1624:smale01apj}.
Though both missions have harder energy range detectors (RXTE--HEXTE detector and 
Beppo-Sax--PDS detector), the diminishing sensitivity and short observation durations
have not yielded effective analysis of light curves and spectra above 20 keV.
This in the end yields spectral parameters that vary according to the detector and mission
regardless of intrinsic source variability.
The ESA's $\gamma$-ray observatory INTEGRAL
will allow for the determination of the widest
energy band observed for dipping LMXB sources with the highest
possible sensitivity above 20 keV. The energy
range of the data ($\sim$ 15--500 keV) provided by the INTEGRAL
observatory ISGRI detector supported by the lower energy JEM-X data 
should reveal the best parameters at high S/N that 
describe the broad-band spectrum of the dipping LMXBs together
with the effects of scattering at high energies.
Furthermore, we extend the investigation of the role of photo-ionized
plasmas via application of warmabsorber models 
in dipping LMXBs beyond the 10~keV upper energy limit of pervious studies 
using data from INTEGRAL observatory. 

In this paper, we derive light curves and spectra for four dipping LMXB
systems, \nineteen, \thirteen, \fouru, and \bigdip.  We
discuss the dipping activity visible in the light curves and
describe the broad-band time-averaged spectral characteristics of
each source. We compare the dip and persistent (non-dip) spectra
of the sources.
We discuss and elaborate on the results of the spectral fits and compare
these with previous findings in the light of the present physical
models of accretion disk coronae and photo-ionized absorbers.

\subsection{Source properties}
\label{sect:prop}

The properties of the four dip sources studied here (\nineteen,
\thirteen, \fouru\ and \bigdip) are summarized in
Table~\ref{tab:properties}.

\nineteen\ is the dip source with the shortest period known of
0.83~hr \citep{1916:walter82apjl}. The presence of dips and the
lack of X-ray eclipses from the companion star indicate that the
system is viewed at an inclination angle, $i$, of
$\sim$60--80$\degmark$ \citep{frank87aa,1916:smale88mn}. By
combining archival and R-XTE observations, \cite{1916:chou01apj}
were able to derive a precise dip recurrence interval of
0.83351411(25)~hr. The optical counterpart shows a
modulation with a period 1\% longer \citep[][ and references
therein]{1916:callanan95pasj}. The presence of superhumps
\citep{1916:schmidtke88aj, 1916:retter02mn}
have been proposed to account for this
difference. \nineteen\ has shown a 270~Hz highly coherent
oscillation during a type~I X-ray burst indicating that the
neutron star is spinning rapidly with a period of 3.7~ms
\citep{1916:galloway01apjl}. The source has also shown
quasi-periodic oscillations (QPOs) at various frequencies ranging
from $\sim$0.2 to 1300~Hz \citep{1916:boirin00aa}. The BeppoSAX
0.2--200~keV non-dip (or "persistent") spectrum may be modeled
using an absorbed blackbody with a temperature, $kT$, of $1.62 \pm
0.05$~keV and an absorbed cut-off power-law with a photon index,
$\Gamma$, of $1.61 \pm 0.01$ and a relatively high cut-off energy,
\ecut, of $80 \pm 10$~keV \citep{1916:church98aa}. Narrow
absorption features due to Fe XXV and FeXXVI in the
XMM-Newton spectrum were discovered by \citet{1916:boirin04aa} which
was interpreted as the existence of a warm absorber in the system. On the
other hand, for  \nineteen\ progressive covering gives a good description of the
spectral evolution in dipping from both ASCA and SAX
observations, as well \citep{1916:church97apj, 1916:church98aa}.

\thirteen\ exhibits X-ray bursts and irregular intensity dips that
repeat every 2.94~hr. The source was first detected by {\it Uhuru}
and {\it Ariel V} \citep{forman78apj, warwick81mn} and the dips
and bursts were discovered using EXOSAT \citep{1323:vdk85ssr,
1323:parmar89apj}. During the dips, which typically last for 30\%
of the orbital cycle, the 1--8 keV intensity varies irregularly
with a minimum of $\sim$50\% of the average value outside of the
dips. The presence of periodic dips and the absence of X-ray
eclipses indicates that the source is also viewed at $i = $
60--80$\degmark$ \citep{frank87aa}. The dip recurrence interval is
not so precisely known, with the best measurement being that of
\citet{1323:balucinska99aa} of 2.94(2)~hr. The source
exhibits $\sim$1~Hz QPOs which have been attributed to a disk
instability \citep{1323:jonker99apjl}. The BeppoSAX 1.0--150~keV
spectrum of the \thirteen\ persistent emission can be modeled by a
cutoff power-law with $\Gamma$ = $1.48 \pm 0.01$ and \ecut\ =
$44.1 \, ^{+5.1} _{-4.4}$~keV together with a blackbody with $kT$
= $1.77 \pm 0.25$~keV \citep{1323:balucinska99aa}.
\citet{1323:boirin05aa} examined the changes in the properties of
the Fe XXV and Fe XXVI absorption features during persistent
and dipping intervals. They found evidence for the presence of
less-ionized material in the line of sight during dips supporting the
warm absorber scenario. \citet{church:mnras05} also explains
the XMM observations of this source using the progressive covering
model where they attribute the absorption features to an existing
accretion disk corona.

\fouru\ is located in the globular cluster NGC\thinspace6441. The
source exhibited 3 apparently energy independent intensity dips
separated by 5.0(5)~hr, as well as flaring and bursting
behavior \citep{1746:parmar89aa}. The dips are shallow (25\% of
the continuum intensity) and largely energy independent in the
usable energy range of the R-XTE Proportional Counter Array
(2.1--16~keV) \citep{1746:balucinska04mn}. The best estimate of
the dip period is $5.16 \pm 0.01$~hr \citep{1746:balucinska04mn}.
The BeppoSAX 0.3--30~keV spectrum of \fouru\ is unusually soft and
is dominated by a disk blackbody with $kT$ = $2.82 \pm 0.04$~keV,
together with a cutoff power-law with $\Gamma$ = $-0.32 \pm 0.80$
and \ecut\ of $0.90 \pm 0.26$~keV \citep{1746:parmar99aa}.

\bigdip\ exhibits irregular dips in X-ray intensity that repeat
every orbital cycle of 21~hr \citep{1624:jones89esaSP}. This long
orbital period means that as well as being the most luminous dip
source, \bigdip\ also has the largest stellar separation and is
often referred to as the "Big Dipper". The \bigdip\ X-ray light
curve exhibits prominent flares at energies $>$15~keV and dipping
is evident at energies up to 6~keV \citep{1624:smale01apj}. These
authors determine a dip recurrence interval of 20.8778(3)
~hr. In contrast to the other dip sources studied, no X-ray
bursts have been observed from \bigdip\
\citep[e.g.,][]{1624:balucinska00aa, 1624:smale01apj}. The
1--100~keV BeppoSAX persistent spectrum of \bigdip\ may be
modelled using a blackbody with $kT$ = $1.31 \pm 0.07$~keV and a
cutoff power-law with $\Gamma$ = $2.0 \, ^{+0.5} _{-0.7}$ and an
\ecut\ of $\sim$12~keV \citep{1624:balucinska00aa}. Narrow
absorption line features identified with Fe XXV and Fe XXVI
were discovered from \bigdip\ using XMM-Newton by
\citet{1624:parmar02aa}.

\section{Observations and data reduction}
\label{sect:obs}

The INTEGRAL \citep{winkler03aa} payload consists of two $\gamma$-ray
instruments, one optimized for 15~keV to 10~MeV high-resolution
imaging \citep[IBIS;][]{ubertini03aa} and the other for 18~keV to
8~MeV high-resolution spectroscopy \citep[SPI;][]{vedrenne03aa}.
IBIS provides an angular resolution of $12\arcmin$ full-width at
half-maximum (FWHM), a fully coded field of view (FOV) of $8\fdg
3$ x 8$\degmark$, and a FWHM energy resolution, $E/\Delta E$, of
12 at 100~keV. SPI provides an angular resolution of $2\fdg
5$~FWHM, a fully coded hexagonal FOV of $13\fdg 2$ x $13\fdg 2$ and a FWHM
$E/\Delta E$ of 430 at 1.3~MeV. The extremely broad energy range
of IBIS is covered by two separate detector arrays, ISGRI
(15--1000~keV) and PICsIT (0.175--10~MeV). In addition, the payload includes
X-ray \citep[JEM-X; 3--35~keV;][]{lund03aa} and
optical \citep[OMC; V-band;][]{mas-hesse03aa} monitors. JEM-X has
a fully coded FOV of $4\fdg8$ diameter and an angular resolution
of $3\arcmin$ FWHM. All the instruments are co-aligned and
operated simultaneously. The data used here were obtained from the
INTEGRAL Science Data Centre \citep[ISDC;][]{courvoisier03aa}
archive. For this study, all publically available pointings between
January 2003 and  November 2004 (revolutions 30 to 250)
where one of the selected sources was within
10$\degmark$ of the centre of the FOVs were included.
Table~\ref{tab:obslog} is an observation log.

For this study, data from the ISGRI and JEM-X instruments were
used due to their good sensitivity and broad energy coverage and
processed using the Off-line Scientific Analysis (OSA~5.1)
software provided by the ISDC. This includes pipelines for the
reduction of INTEGRAL data from all four instruments.
ISGRI and JEM-X use coded masks to provide imaging information. This
means that photons from a source within the field of view (FOV)
are distributed over the detector area in a pattern determined by
the position of the source in the FOV. Source positions and
intensities are determined by matching the observed distribution
of counts with those produced by the mask modulation.  
After the standard data pipeline
processing (dead time correction, good time-interval selection,
gain correction, and energy reconstruction), source spectra and
background subtracted light curves were derived using standard
tools in OSA~5.1\ .  The positions of all the detected sources were
fixed at the values given in the ISDC Reference Catalogue (version
23.0) which contains all known high-energy sources
\citep[see][]{ebisawa03aa} in order to derive reliable source
fluxes. In addition, all sources that are at and above 6 sigma detection limit
were extracted simultaneously during the analysis to avoid incorrect photon
attribution between sources in order to map the field 
correctly. The coverage is less for JEM-X due to its smaller FOV compared 
to ISGRI. For further data manipulation we used standard software tools such as
FTOOLS 6.0.4-6.1.1 and XSPEC 12.2.1-12.3.0. Preliminary analysis of light curves and
spectra can be found in \citet{dippers:balman-esa07,dippers:balman-AIPC08}.

\begin{table*}[ht! for all the four sources for all the four sources for all the four sources for all the four 
sources for all the four sources for all the four sources]
\caption{Observation Log. The longer exposure times for ISGRI
compared with JEM-X are due to the larger FOV of this instrument.
Mean count rates are given in the energy range 4.0--20~keV for JEM-X
and 20--200~keV for ISGRI.}
\begin{center}
\begin{tabular}[c]{lcccccc}
\hline \hline\noalign{\smallskip}
LMXB & \multicolumn{2}{c}{Exposure (ks)}  &
\multicolumn{2}{c}{Count Rate (s$^{-1}$)} &\multicolumn{2}{c}{Tstart-Tstop (MJD)}\\
     & JEM-X & ISGRI  &  JEM-X & ISGRI & JEM-X & ISGRI  \\
\noalign{\smallskip\hrule\smallskip}
\nineteen & 293 & 915 & 3.5  & 1.2 & 52952.4-52963.5 & 52708.7-53310.6 \\
\thirteen & 497 & 834 & 0.96 & 0.66 & 53085.7-53263.9 & 52650.7-53263.9 \\
\fouru    & 316 & 626 & 6.4  & 0.43 & 52713.4-53068.1 & 52713.3-53292.7 \\
\bigdip   & 398 & 756 & 9.4  & 0.42  & 52701.8-53062.0 & 52700.7-53265.2  \\
\noalign{\smallskip\hrule\smallskip}
\end{tabular}
\end{center}
\label{tab:obslog}
\end{table*}

\begin{figure}[ht!]
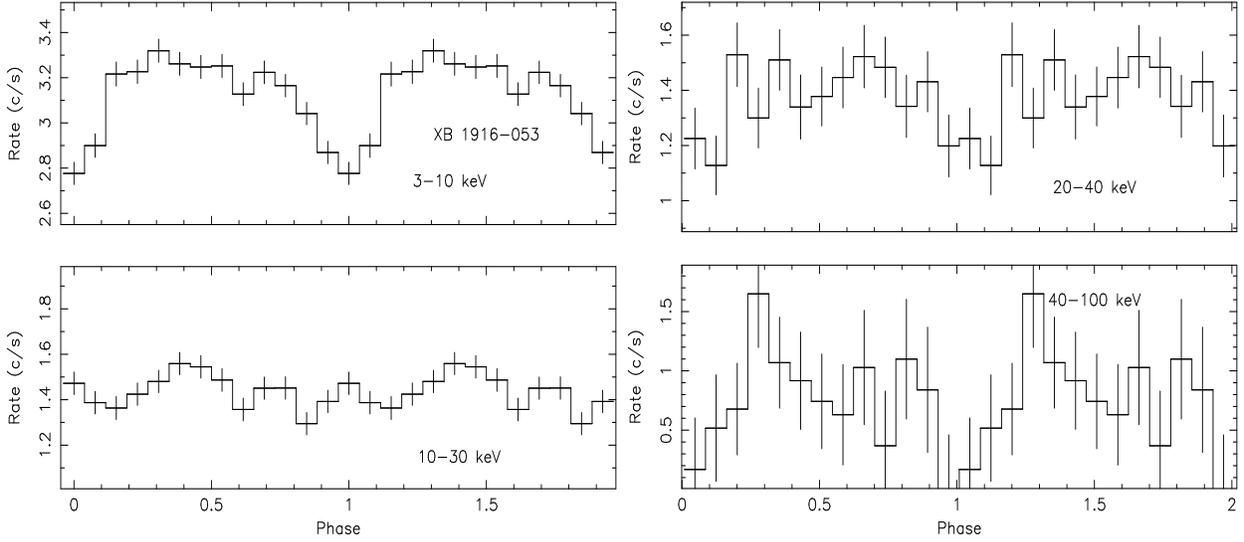

\includegraphics[width=2.8in,height=3.2in,angle=-90]{1916_lcr_j_up.ps}
\includegraphics[width=2.8in,height=3.2in,angle=-90]{1916_lcr_is_up.ps}
\caption{JEM-X (3-30 keV) and ISGRI (20-100 keV)
folded light curves of \nineteen. The energy ranges are
indicated on the panels.
The folding period is given in
Table~\ref{tab:properties}.}
\label{fig:1916foldedlc}
\end{figure}

\begin{figure}[ht!]
\includegraphics[width=2.8in,height=3.2in,angle=-90]{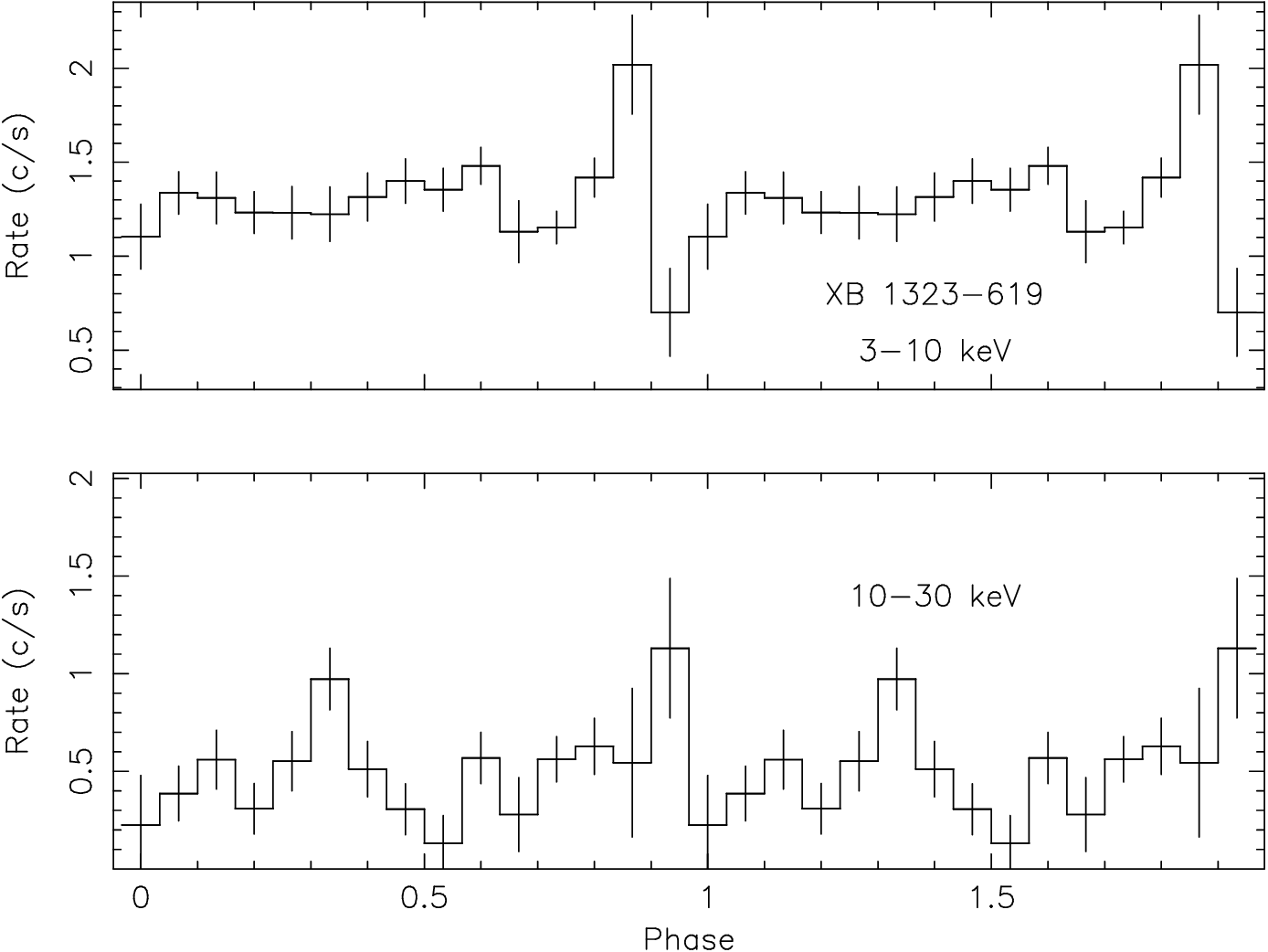} 
\includegraphics[width=1.7in,height=3.2in,angle=-90]{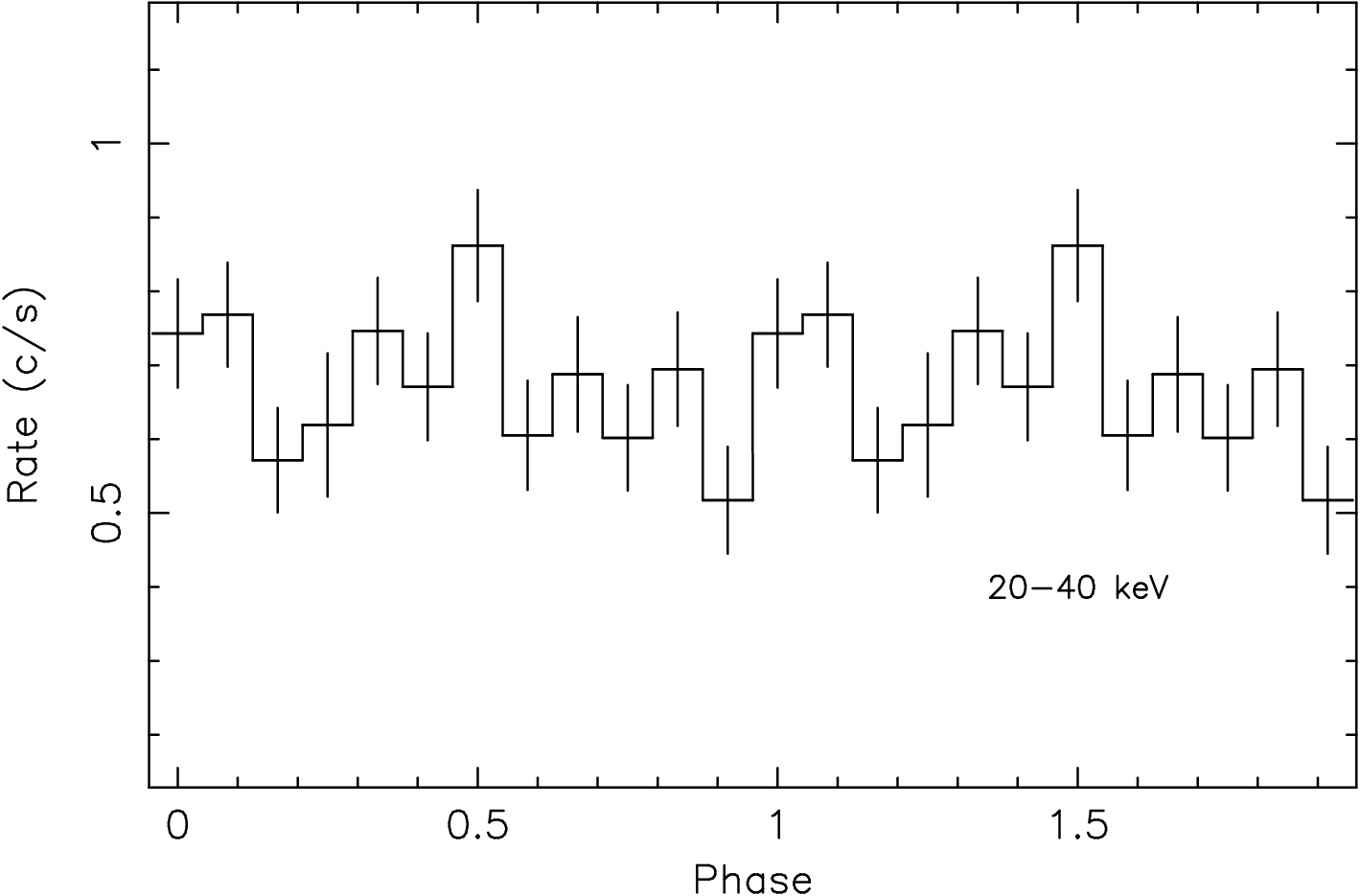}
\caption{JEM-X (3-30 keV) and ISGRI (20-40 keV) folded light curves of \thirteen.
The energy ranges are
indicated on the panels. The folded light curve in the 40--100~keV energy range is not shown
because of the low statistical quality.
The folding period is given in
Table~\ref{tab:properties}.}
\label{fig:1323foldedlc}
\end{figure}

\begin{figure}[ht!]
\includegraphics[width=1.7in,height=3.2in,angle=-90]{1624_lcr_j_up1.ps}
\includegraphics[width=1.7in,height=3.2in,angle=-90]{1624_lcr_is_up2.ps}
\caption{JEM-X (3-10 keV) and ISGRI (20-40 keV) folded light curves of \bigdip. The energy ranges are
indicated on the panels.  The folded light curve in the 40--100~keV energy range is not shown
because of the low statistical quality.
The folding period is given in
Table~\ref{tab:properties}.}
\label{fig:1624foldedlc}
\end{figure}

\begin{figure}[ht!]
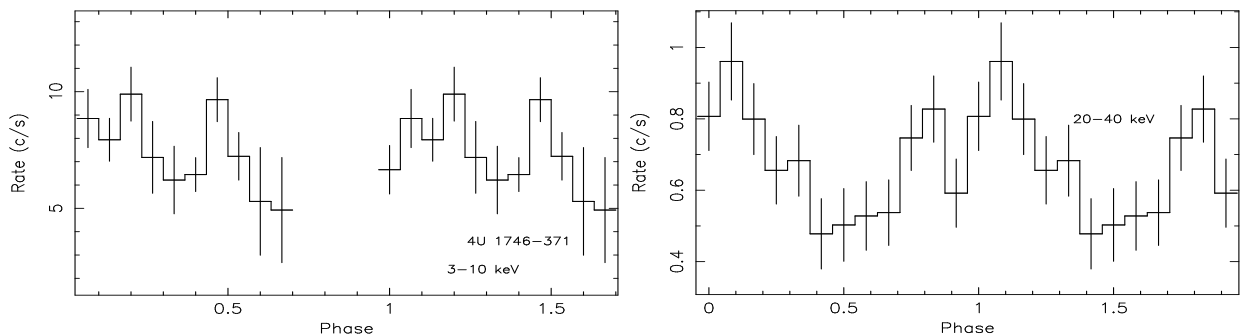

\includegraphics[width=1.7in,height=3.2in,angle=-90]{1746_lcr_j_up2.ps}
\includegraphics[width=1.7in,height=3.2in,angle=-90]{1746_lcr_is_up2.ps}
\caption{JEM-X (3-10 keV) and ISGRI (20-40 keV)
folded light curves of \fouru. The energy ranges are
indicated on the panels. Some of the phases
that are not covered well or show low statistical quality
have been disregarded from the folded light curves.
The folded light curves in the 10--30~keV and 40--100~keV energy ranges are not shown
because of their low statistical quality.
The folding period is given in
Table~\ref{tab:properties}.}
\label{fig:1746foldedlc}
\end{figure}

\section{Results}

\subsection{Light Curves}
\label{subsect:lightcurve}

ISGRI and JEM-X light curves for all the four sources were derived
using the standard OSA~5.1 extraction algorithms. Event times
were corrected to the solar system barycenter
when necessary.
The data were grouped in bins of 10~s (\nineteen), 180~s
(\thirteen), 120~s (\bigdip) and 10~s (\fouru) for the JEM-X
analysis. The grouping time was increased to 240~s (\nineteen),
240~s (\thirteen), 240~s (\bigdip) and 300~s (\fouru) for the
ISGRI analysis in order to increase the signal-to-noise ratio
of the light curves.  A search was made for
any bursting activity since \thirteen, \fouru\ and \nineteen\ are
known X-ray burst sources. For \nineteen, we detected 14
X-ray bursts with JEM-X, one of which is double peaked. For
\thirteen\ some bursts with very low statistical quality
were detected and removed from the data. No bursts were detected from
\fouru\ or \bigdip\ in either the JEM-X or ISGRI data.
Some high intensity data points that appeared in \bigdip\ during
non-dipping intervals
have been removed while constructing spectra. These are most likely
a result of few flaring activities of the source.
For \nineteen\ and \bigdip\ which have well determined dip
periods, the uncertainties on the dip occurrence times over the
observation duration are small compared with their orbital periods so
the light curves from these sources were
folded on the dip periods given in Table~\ref{tab:properties}.
For \thirteen\ and \fouru\ the dip periods are not well
determined, but the error that should be quadratically  added
from binning of the light curves to have enough
signal to noise does not allow
any period error determination better than presently known. We,
therefore, also used the periods given in Table~\ref{tab:properties} for folding.
In addition, the phases of the two light curves are arbitrary since we can not
lock our epoch with the previous ephemerides due to the error in the periods for these two sources.

Overall, we find dipping/modulation of light curves in all sources
(not necessarily in all energy bands) folded on the periods in  Table~\ref{tab:properties}.
The modulation depth normalized to the mean count rate
can be defined as
$\%$ modulation depth $\sim$\ 100$\times$ (max rate - min rate)/(max rate + min rate)/2.
\nineteen\ indicate a modulation of
18$\%$ in the 3--10 keV energy range. 
Above 10 keV the JEM-X observations do not show evidence for any
dipping activity for the source (see Fig.~\ref{fig:1916foldedlc}).
With ISGRI, the 20--40 keV reveals about 25$\%$ modulation depth.
\nineteen\ reveals only
a small modulation of the light curve $<$~15$\%$ at the orbital period in the
40--100 keV energy range (see Fig.~\ref{fig:1916foldedlc}).
For \thirteen, the JEM-X  data show modulation of
40$\%$ in the 3--10 keV energy range for a small range of phases.
It shows no modulation in the higher energies with JEM-X.
The ISGRI, 20--40 keV, data show no variation on the X-ray dip period, as well
%The uncertainty on the period determination of \thirteen\ are 
%large (see Table~\ref{tab:properties}). and thus the almost flat 
%behaviour of the phase folded ISGRI light curve
(see Fig.~\ref{fig:1323foldedlc}).
We note that this could be both because of a physical effect (e.g., precession of the disk)
or due to large error in period determination.
\bigdip\ shows
dipping/modulation in the JEM-X light curve of about 25$\%$
(see Fig.~\ref{fig:1624foldedlc}). The ISGRI light curve of the source has a
modulation of about 40$\%$.
We detect some orbital modulation from \fouru\ in the 3--10~keV (JEM-X)
which is less than 10$\%$.
The ISGRI light curve shows dipping with about 45$\%$ modulation depth
 (see Fig.~\ref{fig:1746foldedlc}). A summary of modulation depths at certain
energies for each source can be found in Table~\ref{tab:modulation}.

\begin{table*}[ht!]
\caption{A summary of modulation depths for the dipping LMXBs analyzed in this work.
The depths are given in percentage modulation normalized to mean count rate 
($\sim$\ 100$\times$ (max rate - min rate)/(max rate + min rate)/2).}
\begin{center}
\begin{tabular}[c]{lccc}
\hline \hline\noalign{\smallskip}
LMXB & 3-10 keV & 20-40 keV & 40-100 keV \\
\noalign{\smallskip\hrule\smallskip}\\
\nineteen & 18$\%$ & 25$\%$ & $<$15$\%$  \\
\thirteen & 40$\%$ &  $<$14$\%$ & -- \\
\fouru    & $<$10$\%$ & 45$\%$ & --  \\
\bigdip   & 25$\%$ & 40$\%$ & --  \\
\noalign{\smallskip\hrule\smallskip}
\end{tabular}
\end{center}
\label{tab:modulation}
\end{table*}

\begin{figure*}[ht!]
\centerline{
%\hspace{0.1cm}
\includegraphics[width=2.0in,height=3.0in,angle=-90]{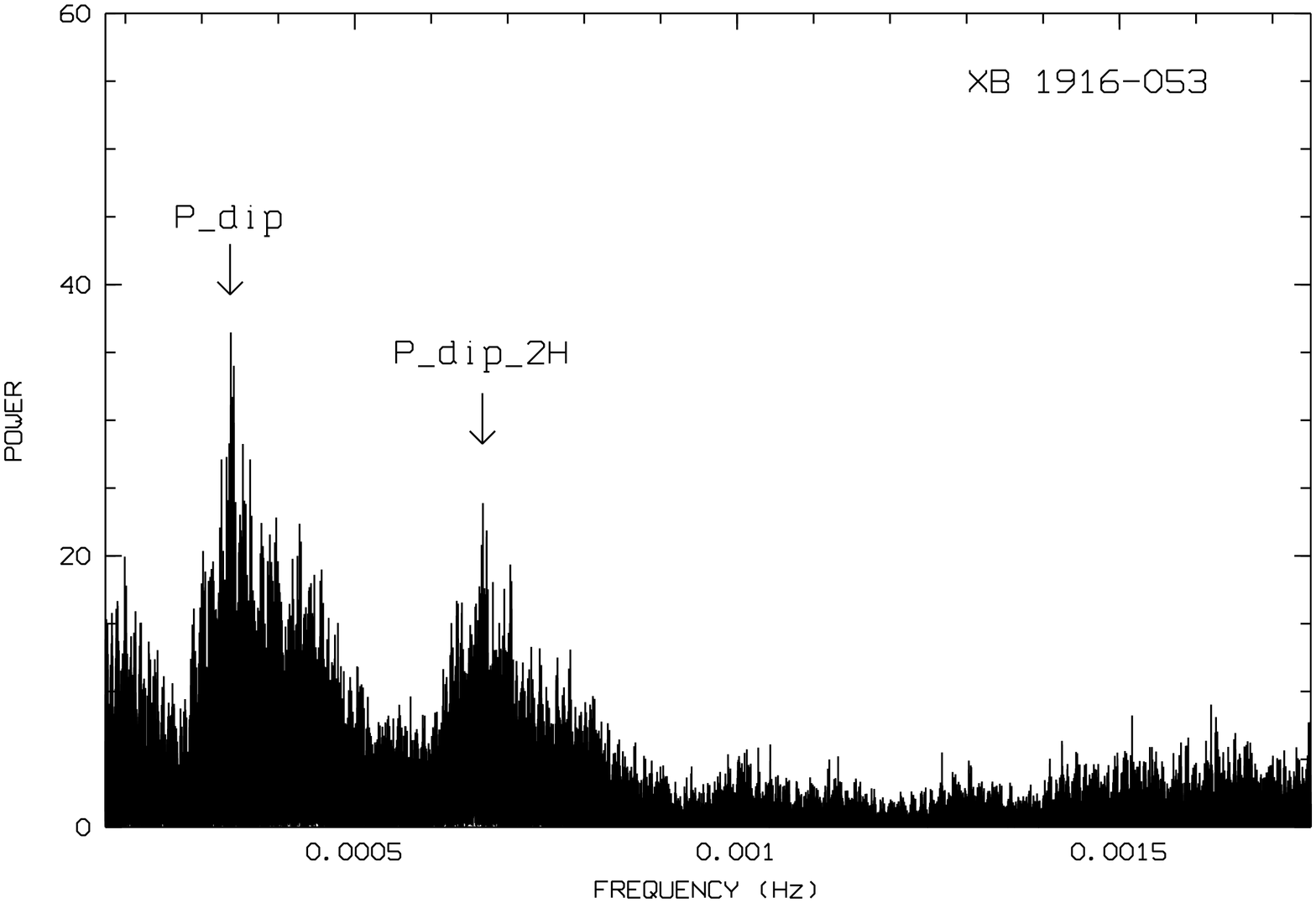}
\hspace{1.5cm}
\includegraphics[width=2.0in,height=3.0in,angle=-90]{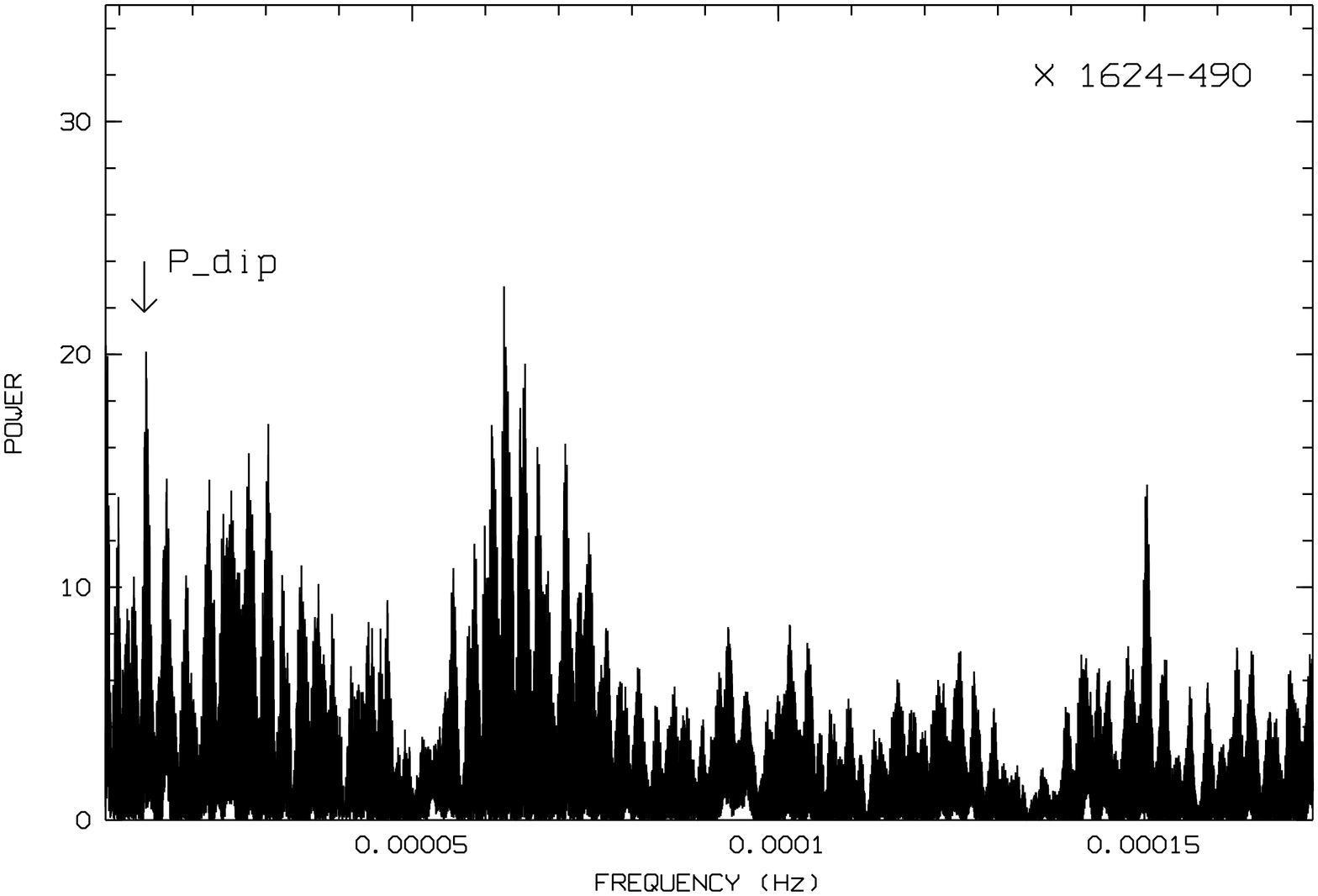}}
\centerline{
%\includegraphics[width=2.0in,height=3.0in,angle=-90]{1624_lpw_up.ps}
%\hspace{1.5cm}
%\includegraphics[width=2.0in,height=3.0in,angle=-90]{1746_lpw_up.ps}}
\includegraphics[width=2.0in,height=3.0in,angle=-90]{1746_hump.ps}}
\caption{Top panels show the power spectra of \nineteen\ and \bigdip\ between 3--10 keV derived using
the Scargle algorithm (i.e., Scargle Periodograms).
Source names are given on each panel. The X-ray dip periods are labeled
on the panels and 2H refers to the second harmonic of the period. See text for 3$\sigma$
detection limits for \nineteen\ and \bigdip. Bottom panel is the average 20--40 keV power spectra of \fouru.
The expected (white) noise level is subtracted and the lowest level on the
y axis corresponds to the 3$\sigma$ detection limit for \fouru.}

\label{fig:JEMXpw}
\end{figure*}

We performed Fourier analysis of the
time series obtained from the JEM-X and ISGRI data in order to recover the known frequency
of orbital modulations, search for new periodic
signals, and detect any
red-noise features in the low frequency band for \nineteen, \bigdip, and \fouru.
The Scargle Algorithm \citep{scargle82apj, scargle:press92apj}
has been applied in the 3--10 keV
range to analyze the JEM-X time series
data. Top panels of Figure~\ref{fig:JEMXpw} show Scargle periodograms
(no power spectra are averaged ).
The X-ray dip periods are noted on each panel of the figure (The periods are listed on Table~\ref{tab:properties}. 
The high peaks in these panels
other than the noted periods are false features of high noise.
The detection limit of a period at the 3$\sigma$ confidence level (99$\%$) is
at a power level of 16.8 for \bigdip\ and 15.3 for \nineteen\ and \fouru\ \citep{scargle82apj}.
Therefore, the signals (at the dip periods) are at 3$\sigma$
level for \nineteen\ and \bigdip, but the
signal is insignificant for \fouru and thus, the Scargle periodogram has been omitted.
We note that the length of the time series used
for both \bigdip\ and  \nineteen\ does not yield better errors than the published ones, thus
we avoid comparison of any derived periodicity with the previous detections. However,
Figure~\ref{fig:JEMXpw} confirms our folded light curves.
In order to correct for the effects of windowing and sampling functions on power spectra,
synthetic constant light curves were created and the actual power spectra were checked
against the synthetic power spectra for false features.
In addition, we did not perform prewhitenning or detrending of the light curves.

We also performed power spectral analysis of
the 20--40 keV ISGRI light curves in a similar way to the JEM-X lightcurves using Scargle
algorithm. However, the periodograms yielded no significant peaks in the power spectra of the
three sources. In order to find the origin of the modulations detected in the ISGRI light curves
we calculated stacked and averaged power spectra using standard Leahy normalization \citep{leahy83apj}.
We have averaged 108, 199, and 11 power spectra for \nineteen,
\bigdip\ and \fouru\ with 2048, 1024, and 16384 frequency bins in each PSD, respectively.
We detect a wide red-noise hump around/in the location of the
orbital/dip frequencies in these sources over 3$\sigma$ (99$\%$ Confidence level) detection level
and particularly a read noise peak at the location of the X-ray
dip period of \fouru (see Fig.~\ref{fig:JEMXpw} lower/bottom panel).
We do not detect a particular
flattening of the red-noise in the 20--40 keV energy band in the frequency range of the
orbital/dip periods which should be expected from disk fed systems
\citep{belloni-psaltis-vdk:apj02}.
We think that the modulations detected above the 20 keV range in  Figs.1--4
are affected from these red noise humps, but the peak in \fouru is clear 
and it shows the highest percent modulation in the 20-40 keV range.
%We attribute it to
%variability resulting from Compton up-scatering in the
%accretion disk coronae (ADC) or due to scattering off of a structure fixed 
%in the outer disk
%like the bulge at the stream impact zone. As the soft photons get absorbed 
%by this bulge,
%the higher energy photons may be preferentially scattered.

\begin{figure*}[ht!]
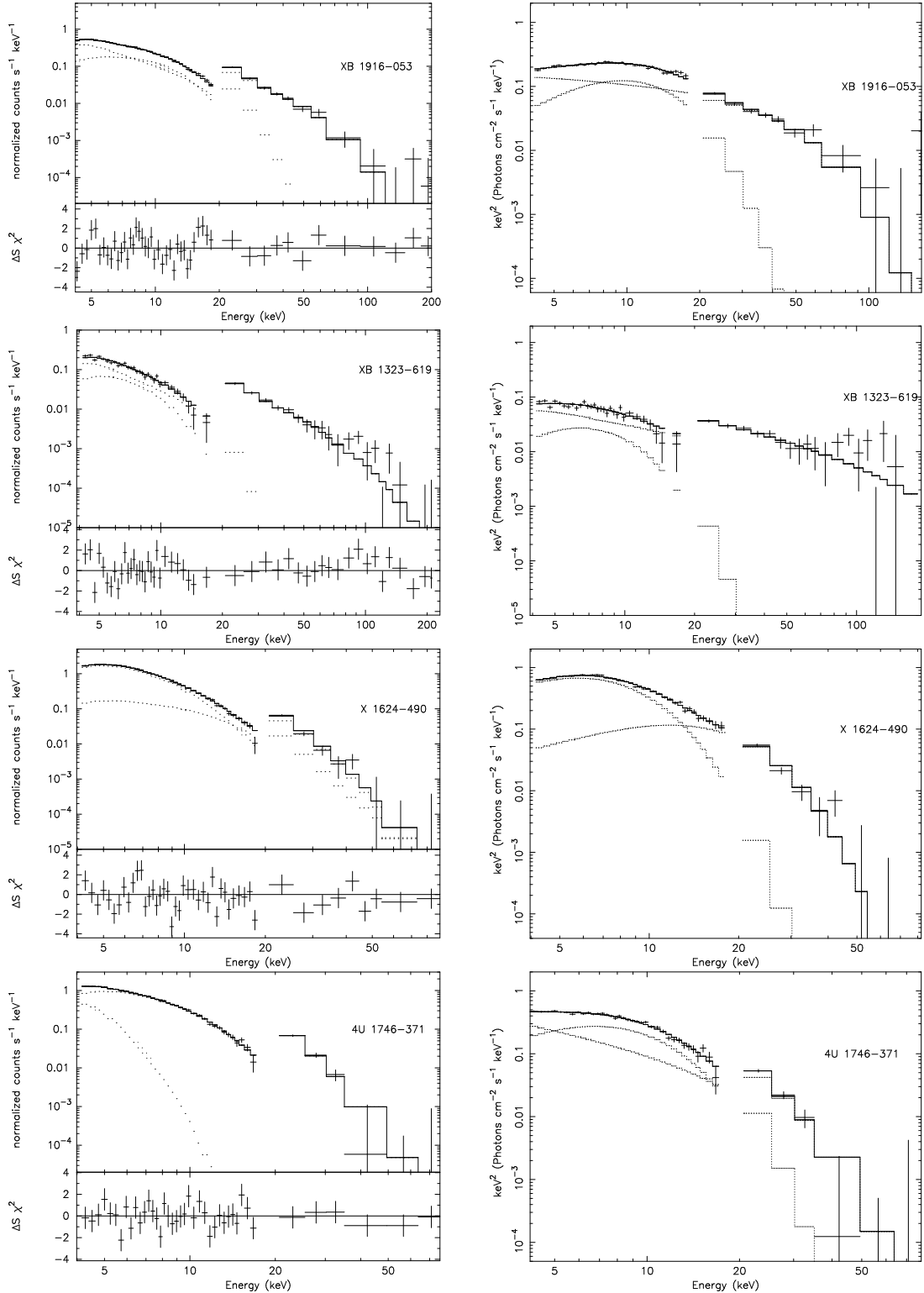

\centerline{
\includegraphics[width=2.0in,height=2.7in,angle=-90]{1916_sp_up1.ps}
\hspace{0.5cm}
\includegraphics[width=2.0in,height=2.7in,angle=-90]{1916_uf_up1.ps}}
\centerline{
\includegraphics[width=2.0in,height=2.7in,angle=-90]{1323_sp_up1.ps}
\hspace{0.5cm}
\includegraphics[width=2.0in,height=2.7in,angle=-90]{1323_uf_up1.ps}}
\centerline{
\includegraphics[width=2.0in,height=2.7in,angle=-90]{1624_sp_up1.ps}
\hspace{0.5cm}
\includegraphics[width=2.0in,height=2.7in,angle=-90]{1624_uf_up1.ps}}
\centerline{
\includegraphics[width=2.0in,height=2.7in,angle=-90]{1746_sp_up11.ps}
\hspace{0.5cm}
\includegraphics[width=2.0in,height=2.7in,angle=-90]{1746_uf_up1.ps}}

\caption{Simultaneous 4--200 keV fits to the JEM-X and ISGRI
spectra of the four dipping LMXBs (4--20 keV; JEM-X and 20--200~keV; ISGRI).
The crosses indicate the data, the solid lines show the summed model and the dotted
lines show the individual models. The second panel under each count rate spectra
display the residuals in standard deviations.  Next to each count rate spectrum
an unfolded spectrum in E$^2$F is plotted to show the contribution of
each component of the spectrum.} \label{fig:spectra}
\end{figure*}

\subsection{Spectra}
\label{subsect:spectra}

The JEM-X and ISGRI data presented in Table~\ref{tab:obslog} were
used to derive a time-averaged broad-band spectrum of each of the
four LMXBs studied in this paper.
%The spectra that showed divergent behaviour have been excluded,
%particularly containning burst-like episodes. 
This data set provides a unique test
that could only be performed with INTEGRAL
since it is the only high-energy mission that frequently observes the
galactic bulge region with large field of view instruments.
%Figure~\ref{fig:lc} shows that 
During the time span of our
observations, we do not detect any prominent high- or low-states
of any of the LMXBs in our sample.

We first performed spectral analysis on combined JEM-X and ISGRI spectra
using XSPEC version 12.2.1 and 12.3.0 \citep{arnaud96conf}. A constant factor
was included in the spectral fitting to allow for a normalization
uncertainty between the two instruments. This factor was
constrained to be within the expected range (i.e., 0.6--0.7) as
 (see \citet{ALLcal:Kirsch05SPIE}) as the parameter was thawed during
the fitting pocedure.
Spectral uncertainties are given at 90\%
confidence level ($\Delta$\chisq = 2.71 for a single parameter) and upper limits at 95\%
confidence. We regrouped the JEM-X energy channels (1--256) by a
factor of 3--4 to increase the signal-to-noise in the spectra. We
used linearly sampled 40 channel response matrices
for the ISGRI data where the data were similarly
regrouped by a factor of 4--6. After the regrouping process,
energy channels that have low signal-to-noise ratio have also been ignored. To account for
systematic effects we added quadratically a 1\% uncertainty to
each spectral bin as recommended in the OSA~5.1 cookbook.
We note that the addition of 1-2\% uncertainty for statistical
concerns only {\it reduces} the $\chi^2_{v}$\ values,
but does not affect the best fit parameters or parameter ranges.

We fitted composite models to the broad-band spectra
consisting of a disk blackbody \citep{mitsuda84pasj,
makishima86apj, 1608:mitsuda89pasj} or a blackbody, together with either a
cutoff power-law or {\tt CompTT} model \citep{tita94apj, hua:apj95}.
Table~\ref{tab:spectra} shows the best-fit spectral parameters and
their uncertainty ranges at 90$\%$ confidence level for each
composite model. The disk blackbody or blackbody model was used to
account for the emission from the disk and/or neutron star. The
emission from an extended scattering/Comptonizing region was accounted for by
the XSPEC model {\tt CompTT}, a more physical model compared with
the generic cut-off power-law model that only involves a simple
power law decay with a high energy cut-off depending on the energy
of the scattering electrons. During the fitting procedure, we also included
the {\tt wabs} XSPEC model to account for the  neutral hydrogen  absorption.
These values  were fixed at the values
derived from the XMM-Newton data \citep{1323:boirin05aa, ionabs:diaz06aa}
(also compared with the N$_H$ values derived from the Beppo-SAX and R-XTE data
to make sure all of them are similar).
Figure~\ref{fig:spectra} displays the fitted broad-band spectra of the four LMXBs.

We also applied a single model
to fit the four broad-band spectra. We used four models; the cut-off power
law, {\tt CompTT}, blackbody, or the disk blackbody for the analysis  and
found that none of the four LMXB spectra are consistent with a single emission model
yielding reduced $\chi^2$ values $>$2.5 and/or nonphysical
parameters except for \thirteen. A fit of a power-law model with
absorption to the spectra of \thirteen\ yields an N$_H$ of 5.5$_{-2.9}^{+3.0}$$\times$
$10^{22}$~cm$^{-2}$ with a power law photon index of 2.9$_{-0.09}^{+0.08}$ and a normalisation
of 0.6$_{-0.2}^{+0.15}$ photons~keV$^{-1}$cm$^{-2}$s$^{-1}$ at 1 keV ($\chi^2_{v}$~=~1.5).
We also fitted the 4-200 keV spectra with a double blackbody or double disk-blackbody model.
The fits resulted in reduced $\chi^2$ values $>$2.0 and/or nonphysical parameters for all
sources.

The X-ray spectrum of \bigdip\ and \thirteen\ as been modeled using a halo model because of the
extensive dust in the vicinity of the sources particularly of \bigdip\ \citep{1624:xiang07apj,
1624:smale01apj,ionabs:diaz06aa,
1624:balucinska00aa,1323:barnard01aa}.
The radial intensity profiles of a source is expected to have excess over the point spread 
function due to the level of dust scattering at low energies mainly below 5-6 keV.
We also applied the $dust$ model in XSPEC to
the INTEGRAL spectra along with the two-component model of emission. 
This multiplicative model modifies a spectrum due to scattering off dust on the line-of-sight. 
The model assumes that the scattered flux goes into a uniform disk whose size has a 1/E 
dependence and whose total flux has a 1/E$^2$ dependence.
The spectral results in  Table~\ref{tab:spectra} do not change with the inclusion of the 
multiplicative model. This is expected
given the sensitivity, the large scale of the point spread function of INTEGRAL detectors and 
the energy range we are applying the model which is above 4 keV. We note that the detected
blackbody temperatures (for the the two sources) are the same as the Beppo-Sax results including
the dust contribution which means that the INTEGRAL spectra is not affceted by 
dust scattering component.

\begin{table*}[ht]
\caption{Best-fit spectral parameters obtained using a cutoff
power-law or {\tt compTT} model together with either a disk blackbody or blackbody
component. \nh\ is the absorbing column in units of
$10^{22}$~atom~cm$^{-2}$, $kT$
is the blackbody or disk blackbody temperature in keV,
$K{\rm_{D}}$ and $K{\rm _{B}}$ are the normalisations of the disk
blackbody and blackbody models,
$\Gamma$ is the power law photon index, \ecut\ is the cutoff
energy in keV, $K{\rm _{cut}}$ is the model normalisation photon~cm$^{-2}$~s$^{-1}$~keV$^{-1}$ at 1 keV,
$F$ is the 4--200~keV flux in units of 10$^{-10}$ erg cm$^{-2}$ $s^{-1}$ and $L$ is the
4--200~keV source luminosities at the distances given in
Table~\ref{tab:properties} in units of 10$^{36}$ erg $s^{-1}$,
$kT_{\rm 0}$ is the input soft photon (Wien) temperature (keV),
$kT_{\rm P}$ is the plasma temperature (keV) and
$\tau$ is the plasma optical depth (electron scattering opacity).}

{\tiny

\begin{tabular} {lllllllll}
\hline \hline\noalign{\smallskip} \multicolumn{1}{l}{Model} &
\multicolumn{4}{c}{\ Disk Blackbody + Cutoff Power-law } &
\multicolumn{4}{c}{\ Blackbody + Cutoff Power-law }\\
\noalign{\smallskip\hrule\smallskip}
 & \nineteen & \thirteen & \bigdip & \fouru &  \nineteen & \thirteen & \bigdip & \fouru \\
\noalign{\smallskip\hrule\smallskip}
 N${\rm _H}$  &
0.4 ($<$2.9) &
3.5 ($<$3.9) & 2.9$^{+4.5}_{<}$  & 0.4 ($<$1.3) & 0.4 ($<$5.2) & 3.5 ($<$5.0) &
3.5$^{+3.6}_{-1.9}$ & 0.4 ($<$3.1)  \\

$kT$  &

3.5$^{+0.2}_{-0.2}$  & 2.3$^{+0.4}_{-0.2}$ & 3.4$^{+0.4}_{-0.3}$ &  2.3$^{+0.1}_{-0.2}$ & 2.4$^{+0.1}_{-0.1}$ &
 1.7$^{+0.3}_{-0.2}$ & 1.4$^{+0.1}_{-0.1}$  & 1.7$^{+0.1}_{-0.1}$ \\

$K{\rm _{D}}$  & 0.14$^{+0.04}_{-0.03}$ & 0.5$^{+0.9}_{-0.4}$
& 0.3$^{+0.3}_{-0.2}$ & 7.4$^{+3.1}_{-2.4}$ \\

$K{\rm _{B}}$  &
   &   & & & 0.004$^{+0.001}_{-0.001}$  &
0.0018$^{+0.0014}_{-0.0011}$ & 0.02$^{+0.008}_{-0.005}$ & 0.019$^{+0.003}_{-0.005}$ \\

$\Gamma$ &
 1.7$^{+0.3}_{-0.3}$ & 2.9$^{+0.2}_{-0.2}$ & -2.0$^{+0.5}_{-0.7}$ & 2.4$^{+0.3}_{-0.2}$ & 2.1$^{+0.3}_{-0.3}$ & 3.0$^{+0.1}_{-0.2}$ & 0.2$^{+0.7}_{-1.5}$ &
2.2$^{+0.7}_{-0.8}$ \\

\ecut\ & 26.7$^{+5.2}_{-3.4}$ & 199.30$^{<}_{-90.00}$ &
1.3$^{+0.2}_{-0.1}$
 & 9.4$^{+2.0}_{-1.2}$ & 26.0$^{+7.5}_{-5.0}$  & 500.00$^{<}_{-390.00}$ & 4.1$^{+0.5}_{-0.5}$ &
6.74$^{+4.5}_{-2.2}$ \\

$K{\rm _{cut}}$  &
 0.05$^{+0.03}_{-0.02}$  & 0.6$^{+0.4}_{-0.3}$ & 0.07$^{+0.04}_{-0.02}$ & 0.8$^{+0.3}_{-0.3}$ & 0.2$^{+0.1}_{-0.1}$ & 0.9$^{+0.6}_{-0.4}$ &  0.044$^{+0.013}_{-0.013}$ & 2.2$^{+2.3}_{-1.2}$ \\

$F$ & 5.7 & 1.8 & 20.7 & 9.5 & 6.9 & 1.9 & 12.4 & 8.5 \\

$L$ & 5.6 & 2.2 & 56.2 & 13.1 & 6.7 & 2.3 & 33.7 & 11.8 \\

$\chi^2_{v}$\ (dof) &
 1.3 (42) &
1.2 (47) & 1.34 (41) & 1.1 (34) & 1.4 (42) &
1.2 (42)  & 1.4 (40) & 1.14 (36)   \\

\noalign{\smallskip\hrule\smallskip}
\multicolumn{1}{l}{\ Model}
& \multicolumn{4}{c}{\ Disk Blackbody + CompTT }   &
\multicolumn{4}{c}{\ Blackbody + CompTT }  \\
\noalign{\smallskip\hrule\smallskip}
 & \nineteen & \thirteen & \bigdip & \fouru &  \nineteen & \thirteen & \bigdip & \fouru \\
\noalign{\smallskip\hrule\smallskip}

 $N{\rm _H}$  &
 0.4 ($<1.2$) & 3.5 ($<$6.4) & 3.0$^{+1.6}_{-1.6}$ & 0.4 ($<$1.3) & 0.4 ($<$3.1) & 3.5 ($<$3.6) & 3.1$^{+0.1}_{-0.1}$  & 0.4 ($<$12.0)   \\

 $kT$  &
3.6$^{+0.2}_{-0.2}$  & 2.3$^{+0.4}_{-0.3}$  & 3.4$^{+0.4}_{-0.2}$ & 2.3$^{+0.1}_{-0.2}$
 & 2.44$^{+0.10}_{-0.10}$ & 1.7$^{+0.3}_{-0.2}$ & 1.4$^{+0.1}_{-0.1}$ & 0.6$^{+0.1}_{-0.1}$ \\

 $K{\rm _{D}}$  &
0.14$^{+0.04}_{-0.05}$  & 0.5$^{+0.7}_{-0.4}$  & 0.3$^{+0.1}_{-0.3}$ & 7.5$^{+2.9}_{-1.8}$ & & & & \\

 $K{\rm _{B}}$  &
  &   &  &  & 0.0048$^{+0.0003}_{-0.0002}$  & 0.0014$^{+0.0008}_{-0.0007}$   & 0.025$^{+0.002}_{-0.002}$  & 0.04$^{+0.02}_{-0.01}$ \\

 $kT{\rm _0}$   &
0.7$^{+0.3}_{-0.6}$ & 0.4$^{+0.3}_{-0.3}$  & 1.3$^{+2.2}_{-0.2}$ & 0.4$^{+0.8}_{-0.3}$ & 0.7$^{+0.1}_{-0.2}$  & 0.4$^{+0.2}_{-0.3}$ & 2.5$^{+0.3}_{-0.2}$ & 1.6$^{+0.1}_{-0.1}$ \\

 $kT{\rm _{P}}$  &
 12.3$^{+5.2}_{-2.9}$  & 239.5$^{+34.4}_{-47.6}$  & 11.9$^{+3.4}_{-3.1}$ &  4.8$^{+7.8}_{-1.7}$ & 18.8$^{+10.9}_{-6.0}$  & 196.4$^{+28.5}_{-18.8}$ & 9.2$^{+2.5}_{-1.9}$ & 4.9$^{+5.6}_{-1.9}$ \\

 ${\rm {\tau}_{P}}$\  &
2.1$^{+0.8}_{-1.0}$  & 0.003$^{+0.002}_{-0.002}$ & 0.0038$^{+0.030}_{-0.0030}$ & 2.6$^{+3.7}_{-2.4}$ & 0.9$^{+0.7}_{-0.6}$  & 0.006$^{+0.001}_{-0.004}$ & 0.22$^{+4.21}_{-0.20}$ & 1.1$^{+2.3}_{-0.9}$  \\

 $K{\rm _{C}}$  &
0.003$^{+0.013}_{-0.002}$ & 0.0010$^{+0.064}_{-0.0007}$  &  0.03$^{+0.1}_{-0.01}$ &  0.10$^{+2.92}_{-0.08}$ & 0.005$^{+0.003}_{-0.003}$ & 0.0008$^{+0.0120}_{-0.0005}$ & 0.006$^{+0.001}_{-0.001}$ & 0.11$^{+0.10}_{-0.10}$ \\

 $F$ & 5.8 & 1.8 & 12.7 & 9.0 & 5.9 & 3.3 & 21.6 & 11.0 \\

 $L$ & 5.7 & 2.2 & 34.5 & 12.4 & 5.7 & 4.0 & 58.7 & 15.2 \\

$\chi^2_{v}\ (dof) $&
1.3 (41) & 1.2 (42)  & 1.4 (40) & 1.3 (35) & 1.4 (41) & 1.3 (42) & 1.4 (40) & 1.1 (35) \\

\noalign{\smallskip\hrule}
%\multicolumn{9}{l}{\ {\bf S1} is \nineteen, {\bf S2} is \thirteen, {\bf S3} is \bigdip, {\bf S4} is \fouru.}
\end{tabular}
}
\label{tab:spectra}
%\end{sideways}
\end{table*}

We also divided the JEM-X and ISGRI data into dipping and non-dipping
episodes by selecting on phase intervals
and created spectra for the LMXBs in our sample. The dip-phase
bins in Figs.~\ref{fig:1916foldedlc}--~\ref{fig:1624foldedlc} (phases 0.9--1.05) were used
together with the periods in Table~\ref{tab:properties}. The extracted dips are
an average over different levels of dipping, but the phase bins approximates the levels closest to the deepest dips \citet{ionabs:diaz06aa, 1323:barnard01aa}.
We were able to recover statisticaly significant and different
dip and non-dip spectra for \nineteen\ and \bigdip.
The analysis of the \fouru\ data did not reveal any difference in the dip
and non-dip spectra. Given the low count rates and the short dip intervals
observed from \thirteen, we could not derive a statistically significant dip spectra with JEM-X
in 15~ksec (dipping time in the given exposure time) and the high energies (ISGRI) show no
variation, hence no dipping activity at the given period
(see Fig.~\ref{fig:1323foldedlc}).

\begin{figure*}[ht!]
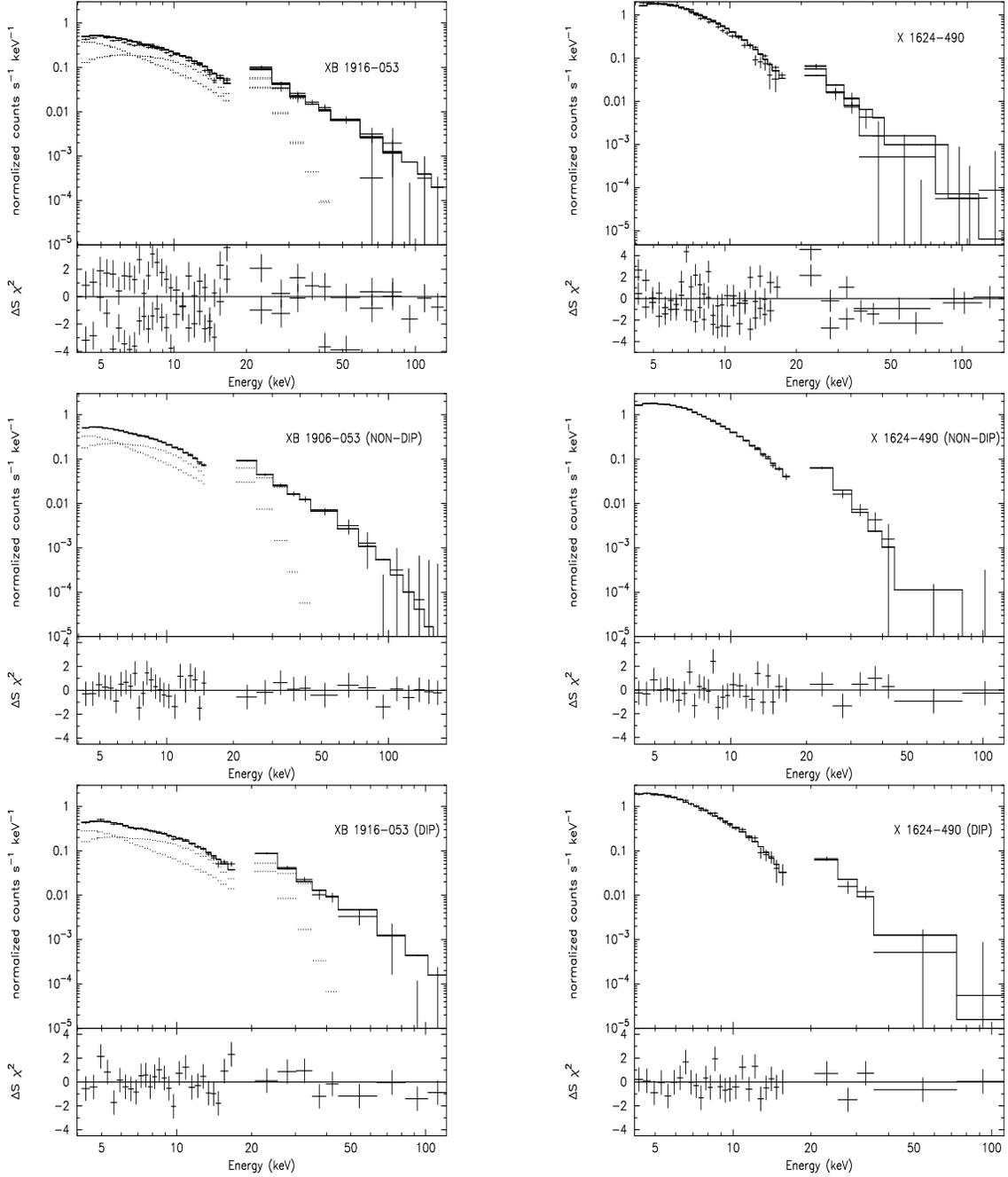

\centerline{
\includegraphics[width=2.3in,height=2.6in,angle=-90]{1916_globalnofit_up.ps}
\hspace{1.5cm}
\includegraphics[width=2.3in,height=2.6in,angle=-90]{1624_globalnofit_up1.ps}}
%\centerline{
%\includegraphics[width=2.3in,height=2.7in,angle=-90]{1916_globalfit.ps}
%\hspace{1.3cm}
%\includegraphics[width=2.0in,height=2.3in,angle=-90]{1624_globalfit.ps}}
\centerline{
\includegraphics[width=2.3in,height=2.6in,angle=-90]{1916_warmabs_per_up.ps}
\hspace{1.5cm}
\includegraphics[width=2.3in,height=2.6in,angle=-90]{1624_per_warm_up.ps}}
\centerline{
\includegraphics[width=2.3in,height=2.6in,angle=-90]{1916_warmabs_dip_up.ps}
\hspace{1.5cm}
\includegraphics[width=2.3in,height=2.6in,angle=-90]{1624_dip_warm_up.ps}}

\caption{Simultaneous 4--200 keV fits to the JEM-X and ISGRI time-average
spectra of \nineteen\ (left) and \bigdip\ (right) (4--20~keV; JEM-X and 20--200~keV; ISGRI).
Crosses indicate the data, the solid lines show the summed model and the dotted
lines show the individual models. The second panel under each count rate spectra
display the residuals in standard deviations.
The top panels are dip and non-dip spectra fitted simultaneously with only the
blackbody and \comptt\ model ({\tt wabs*(bbody+compTT)}).
The middle and bottom panels are non-dip and dip spectra, repectively,
fitted with the composite model including the warm absorber model ({\tt wabs*warmabs*(bbody+compTT)}).
} \label{fig:dipspectra}
\end{figure*}

Fig.~\ref{fig:dipspectra}~ shows the simultaneously fit
dip and non-dip spectra of \nineteen\ and \bigdip\ with only a composite
model of a blackbody and \comptt.
The top right hand panel of Fig.~\ref{fig:dipspectra}
shows the fitted \nineteen\ spectra and the top left panel shows
the \bigdip\ spectra. The composite spectral model was simultaneously fit to the
dip and non-dip spectra of the two sources for a better determination of the differences in the fitted results
(the fitted model was {\tt wabs*(bbody+compTT)}).
The residuals of the top panels of
Fig.~\ref{fig:dipspectra} indicate distinctive deviations from the mean and
the reduced $\chi^2$ values are $>$2.4 in the top panels (the d.o.f. are similar to the ones reported on
Table~\ref{tab:spectra}).
Acceptable fits  were achieved when
two different neutral hydrogen column densities  for dip and non-dip
spectra were assumed along with an additional absorption line at 6.9 keV for
 \nineteen\ (dip spectra) and an emission line for \bigdip\ (non-dip spectra)
in the fitting proccess (additional models of {\tt gabs} and {\tt gauss} were used
for modeling the lines).
The spectral parameters for the composite blackbody and \comptt\ model
of the dip and non-dip spectra for \nineteen\ and \bigdip\ resemble the best fit
values on Table~\ref{tab:spectra} very closely, so we do not
include them in this paragraph. On the other hand, we found a previously detected \citep{1916:boirin04aa} 
absorption line in \nineteen\ at 6.9$^{+0.6}_{-1.8}$ keV.
Addition of the line improves the fluctuation 
of the (fit) residulas from above 2$\sigma$ down to less values in the 
vicinity of 
6-7 keV. We also found a previously detected  emission  line in the
non-dip spectrum of \bigdip\ at 6.8$^{+0.2}_{-0.2}$ keV with an integrated flux  of 2.1$^{+0.9}_{-0.8}$$\times$ 10$^{-3}$ photons s$^{-1}$ cm$^{-2}$. 
Addition of the line improves the fluctuation
of the (fit) residulas from 3$\sigma$ down to less values in the vicinity of 
6-7 keV.

The detection of an increased amount of absorption in dip spectra for 
both sources together with the presence of an absorption line
in \nineteen\ suggests the  existence of photoionized absorbers
solely from the INTEGRAL data. However,
INTEGRAL data, alone, does not account for  existence of photoionized absorbers
in the persistent emission.
Fig.~\ref{fig:dipspectra} middle and bottom panels show the non-dip and
dip spectral fits to both \nineteen\ and \bigdip.
In order to compare with the results from
the XMM-Newton analysis of these two sources with models of photoionized absorbers,
we applied the
warm absorber model, {\tt warmabs} (an additional model in XSPEC), compiled with
XSPEC 12.3.0 to both spectra of \nineteen\ and \bigdip. {\tt Warmabs} models the absorption from a photoionized plasma in the line of sight. Column densities 
of ions (including small cross-sections) are coupled through a photoionization
model. {\tt Warmabs} calculates spectra using stored level populations 
calculated by XSTAR \citep{kallman:apjs01, bautista:apjs01, bautista:apj00} assuming a given continuum
spectrum. For the ionizing flux a photon index of $\Gamma=2$  was 
assumed in this study consistent with the X-ray observations of these systems.
The spectral parameters of {\tt warmabs} 
are an absorption column density of hydrogen of the photoionized plasma, 
the logarithm of the ionization parameter \logxi, and 
the turbulent velocity broadening $\sigma_v$ along with elemental abundances. 
The ionization paramter $\xi$ = L/nR$^2$ where L is the luminosity of the 
ionizing source, n is the  gas density, and R is the distance from the 
ionizing source. The middle panels in
Fig.~\ref{fig:dipspectra}  show
the non-dip and the bottom panels show the dip spectra fitted with the composite model of
a neutral hydrogen absorption, a blackbody and a  \comptt\ including a {\tt warmabs} model
to account for the excessive absorption and/or the absorption line. The
fits performed with the dip and non-dip spectra of \nineteen\ and \bigdip\
yield reduced $\chi^2$ values around 1.0--0.83  (for 36--31 d.o.f.) .
Using the fits to the INTEGRAL data, we derive a \logxi\ of 3.8$^{+0.2}_{-0.4}$ for the persistent
and 2.8$^{+0.2}_{-0.2}$ for the dip spectra of \nineteen.
The absorption column of the
ionized absorber is 4.8$^{+1.3}_{-1.3}$$\times$ 10$^{22}$ cm$^{-2}$
during the dipping episodes and
1.7$^{+1.5}_{-1.5}$$\times$ 10$^{22}$ cm$^{-2}$
during non-dipping intervals. The fits performed with the dip and non-dip spectra of \bigdip\
yield a \logxi\ of 3.4$^{+0.2}_{-0.2}$ for the persistent and 2.6$^{+0.2}_{-0.2}$ for the dip spectra.
The fit results indicate an absorption
column of 4.0$^{+1.9}_{-1.9}$$\times$ 10$^{22}$ cm$^{-2}$ for the ionized absorber in the
persistent and 15.8$^{+1.6}_{-1.6}$$\times$ 10$^{22}$ cm$^{-2}$ in the dipping spectra.
The N$_H$ parameter to the line of sight are kept constant as in 
Table~\ref{tab:spectra} during analysis. The paramter $\sigma_v$ was 
fixed at the values determined from XMM-Newton observations 
\citep{ionabs:diaz06aa};  
fixed at the values of 100 km s$^{-1}$ (dip) and 300 km s$^{-1}$ 
(non-dip) for \bigdip\ and 200 km s$^{-1}$ (dip) and 2000 km s$^{-1}$ (non-dip) for \nineteen.

\section{Discussion}
\label{sect:discussion}

We report the results of archival INTEGRAL observations of
four dipping LMXBs, \nineteen, \thirteen, \fouru\ and \bigdip. The
data presented in Table~\ref{tab:obslog} were used to derive
time-averaged broad-band spectra of the four LMXBs (January 2003--November 2004)
which is a unique capability of the INTEGRAL observatory as the
data analysis is performed by averaging observation science windows over long periods
of time. We, also, investigated the dip and non-dip spectra.
During the time span of  our data, we do not detect a prominent high or low state
of any of the LMXBs in our sample.

The LMXB spectra have been assumed using several different approaches.
One of the approaches, Western model \citep{white88apj}, is a two-component emission model where
the harder X-ray component is the Comptonized emission originating from an ADC in the LMXB spectra
and it has the form of a power-law with an energy cut-off at the limit
of the Comptonizing electrons and only the very bright sources required
an additional blackbody component from the boundary layer between the
accretion disk and the surface of the neutron star.
Another model, Eastern model \citep{1608:mitsuda89pasj},
assumes that photons from the neutron star are Comptonized in a small central region on the disk
and harder component originates from a multi-temperature disk blackbody model.
A descendent model, Birmingham model \citep{1624:church95aa}, takes after the Western model, but assumes
neutron star blackbody for the soft components. High-resolution spectroscopic observations of
LMXBs have shown in some cases the presence of extended X-ray emitting plasmas or ouflows surrounding
the accretion disk. The detection of these components and their kinematic properties has been made possible
by the observation of emission or absorption lines from very highly ionized metals. This also diagnoses
several different emitting zones in LMXBs such as disk atmospheres and coronae
\citep{Jimenez:2002apj,Jimenez:2003apj}.
In this paper, we fitted spectra
assuming several different models and constructed Table~\ref{tab:spectra} to test
the INTEGRAL data given the presumed approaches to derive the
spectral parameters, also, introducing a warm absorber model to model the
absorption over 4--200 keV in the light of recent model construction for dipping LMXBs.

%In general, the 4--200~keV X-ray flux from the time-averaged data
%for individual sources is larger by a factor of 1.0--1.3 compared
%with the findings in the previous studies between 0.5--10 keV, except for
%\thirteen\ (see Table~\ref{tab:properties}) which is about a
%factor of 2 lower compared with the given luminosity in \citet{1323:boirin05aa}.
%See Table~\ref{tab:spectra} for the individual luminosities calculated from the
%INTEGRAL data that range from
%2.2--58.7$\times$$10^{36}$ erg s$^{-1}$ cm$^{-2}$.
%The general discrepancies where the INTEGRAL luminosity is slightly different (e.g., larger)
%can be expected since the energy range of our data is larger and one
%needs to assume statistical uncertainities on spectral parameters involved.

This INTEGRAL study uses better S/N spectra and lightcurves incomparison with other studies, particularly above 20 keV
for the four LMXBs in question. 
None of the RXTE studies of these four sources include usable spectra or light curve above 20 keV. The Beppo-Sax has the
opportunity to extend out to 100 keV using the PDS instrument which has been used in the literature. 
However, the shorter exposures
 15-30 ksec and low count rates, do not allow for a high S/N spectra above 20 keV,
the effective area also falls readily and there are no PDS light curves avaliable in the literature above 20 keV.
In general, for two of the sources \fouru\ and \bigdip, there is
no effective study of the spectra or light curve above 15-20 keV.
For \nineteen\ and \thirteen\ the PDS spectra have been helpful, however the S/N is not favorable to the INTEGRAL study here.
INTEGRAL countrates are three to four times better for above 500 ksec integration time (we used about
1 Msec exposure on each source with ISGRI). As a result ISGRI results in this study provide new information on these
objects and they are supported by the JEM-X detector (to avoid cross-calibration issues that could arise
from using other observatory data) to provide large energy band coverage.

Our spectral results in Table~\ref{tab:spectra} yield blackbody temperatures of 1.7 keV and 1.4 keV,
for \thirteen\ and \bigdip\ which are in excellent agreement with the
BeppoSax results (see~\ref{sect:introduction}) within statistical errors.
On the other hand, \nineteen\ and \fouru\ yields different blackbody temperatures
of 2.3 keV and 0.6 keV, respectively where the BeppoSax results reveal 1.62 keV and
0.5 keV for the same sources (see~\ref{sect:introduction}). The INTEGRAL result
and BeppoSax results for \fouru\ agree within statistical errors.
Moreover, the Wien photon  temperature kT$_0$ (parameter--seed in photons)
varies in a range 0.4--2.5 keV in the four sources (even if we derive similar
blackbody temperatures). These differences could be attributed to
the existence of more than one blackbody component for at least the cases like \nineteen\ and \fouru.
At suffciently high mass accretion rates, the boundary layer and the accretion disk
are both radiating in the optically thick regime with kT 1--2 keV. It is expected on theoretical
grounds and demonstrated that the boundary layer has higher temperature than the accretion disk
yielding two different blackbody temperatures in the X-ray spectra \citep{gilfanov:2003, gogus:2007}.

We find that the normalisation parameter for the disk blackbody model,
proportional to $R_{in}^2 cos\theta$,
yields fairly small real inner disk radii of less than 14.5 km, 5.2 km, and 12.8 km (all at 3$\sigma$ maximum limit)
for \thirteen, \nineteen, and \bigdip,
respectively ($\theta$ is the inclination angle suitably taken between 60$\degmark$--80$\degmark$).
We stress that at 2$\sigma$  limit these radii are less than a neutron star size.
On the other hand, this value is between 27-37 km for \fouru. As a result, a
disk blackbody emission is very unlikely for the three sources except for
\fouru. We note that the 3$\sigma$ maximum limits on apparent inner disk radii
are corrected to real radii according to \citet{kubota:98}.
We caution that normalisation of the disk blackbody
model involve distance in the calculation which results in a slight
uncertainity. Radii change only linearly with distance and only a factor of 2-3 times (in kpc) difference
in distance will effect the general trend discussed above.
In comparison with this, we derived the size in radii of the blackbody emitting regions for
the four sources. We calculate that for \thirteen\ and \nineteen\ the radii is between 0.8-1.02 km.
For \bigdip, this radius is about 9.7-11.1 km (e.g., the size of a neutron star). \fouru\ shows
an emitting region of 37-51 km which is physically inconsistent with emission from the surface of
a neutron star. These ranges above correspond to 3$\sigma$ error ranges of the used parameters.
INTEGRAL results suggets that
\thirteen\ and \nineteen\ accrete to a small region on the surface of the neutron star
and \fouru\ shows only a disk blackbody or boundary layer emission.
The disk blackbody temperature of \fouru  detected in this study is consistent 
with the BeppoSax results as well (see~\ref{sect:introduction}) with only slight difference in 
disk blackbody temperature.
It is difficult to make such deductions for  \bigdip\ using solely INTEGRAL data except that a
disk blackbody emission is inconsistent with the source.
In addition, we note that the normalisation of the blackbody model involve distance in 
the calculation
resulting in some uncertainity.
Radii change only linearly with distance. As a result, a factor of only 10 times difference in distance for the
case of \nineteen\ and \thirteen\
and a factor of only 2-3 difference in the case of \bigdip\ will effect the general trend discussed above.

A comparison with the spectral results obtained from the BeppoSax and R-XTE observations
using the Cut-off
power-law model for the second emission component shows that the time average data yield somewhat different
results for the \ecut\ and the photon index parameters (see
Table~\ref{tab:spectra} for individual results). For \nineteen,
\citet{1916:church98aa} find $\Gamma$ = 1.61$\pm$0.01 and \ecut\ = 80$\pm$10~keV and \citet{Narita:apj03}
find a range of powerlaw indicies between 1.74-2.05 and \ecut\ of $>$ 30 keV (90$\%$ confidence limits). Our results
are consistent with R-XTE observation analyses.
For \bigdip, \citet{1624:balucinska00aa} find
$\Gamma$ = 2.0$\pm$0.06 and \ecut\ $>$12~keV where we find a
photon index $>$ 0.84 and a low \ecut\ value of 1.23--4.6 keV including
uncertainty ranges.
For \fouru, \citet{1746:parmar99aa} find
$\Gamma$ $<$ 0.5 and \ecut\ = 0.90$\pm$0.26~keV where we find $\Gamma$ = 2.18$\pm$0.8
and \ecut\ value of 6.6--11.2~keV. For \thirteen, \citet{1323:balucinska99aa}
find $\Gamma$ = 1.48$\pm$0.01 and \ecut\ = 44$\pm$5.1~keV (BeppoSax results) and \citet{1323:barnard01aa} derive
$\Gamma$ = 1.23$\pm$0.07 and \ecut\ $>$44~keV (R-XTE results). Our results yield
$\Gamma$ = 3.0$\pm$0.2 and an unbound cut-off energy larger than 156 keV. \citet{1323:barnard01aa} results show
an $\Gamma$ = 1.89-1.91 for  \ecut\ $>$ 117, however without inclusion of a blackbody model (90$\%$ confidence limits).
These are discrepancies regardless of error estimates and such differences between
 the \ecut\ and the photon index parameters exists also
between  BeppoSax and R-XTE results.
Apart from the fact that there are possible cross-calibration inconsistencies, and that the sources could be
variable,
this could also be attributed to the Cut-off power-law model being an inadequate model of emission to represent the
data in the harder energies particularly in the ISGRI energy band (e.g., relativistic effects).
This strengthens the fact that the
Comptonizing plasmas in these systems needs to be modeled more properly.

We applied the {\tt CompTT} model in XSPEC
inorder to account for the Compton scattering regions in
these systems and found a range of plasma temperatures of 18.8$^{+10.9}_{-6.0}$ keV for
\nineteen, 196.4$^{+28.5}_{-18.8}$ keV
for \thirteen, 9.2$^{+2.5}_{-1.9}$ keV for \bigdip\ and 4.9$^{+5.6}_{-2.0}$ keV for \fouru\
given the best fit
results of time-averaged data. These temperatures indicate
persistent hot Comptonizing regions, accretion disk coronae (ADCs), in these systems.
%The plasma temperature of
%\thirteen\ is $>$123~keV and the parameter is unconstrained.
%The most probable model for this source is then a power law as the higher energy component.
The optical depth of Compton scattering regions can, also, be derived from the fits,
$\tau$, is in a range 0.9$^{+0.7}_{-0.6}$  for \nineteen,
0.006$^{+0.001}_{-0.004}$ for \thirteen, 0.2$^{+4.2}_{-0.2}$
for \bigdip\ and 1.1$^{+2.3}_{-0.9}$ for \fouru\
which, in addition,  indicates that the Comptonizing
regions are relatively tenuous. \thirteen\ has the lowest value. Using the
model parameter descriptions of {\tt CompTT}, n${\rm
_e}$$\simeq$$\tau^2$/r$_{\rm ad}$$\sigma {\rm _T}$ at high optical
depth \citep{tita94apj, hua:apj95}, we derived maximum limits for the
electron densities in the X-ray emitting regions of the
four systems as: \nineteen,
n${\rm _e}$$<$7.6$\times$$10^{14}$ cm$^{-3}$, \fouru,
n${\rm _e}$$<$1.4$\times$$10^{15}$ cm$^{-3}$, \bigdip,
n${\rm _e}$$<$1.1$\times$$10^{15}$ cm$^{-3}$, and \thirteen,
n${\rm _e}$$<$5.4$\times$$10^{9}$ cm$^{-3}$.
%These values are generally
%lower than the expected for accretion disk coronae that are compact/small
%regions, but for \thirteen\ it is quite low. 
We have assumed
0.25$\times$r$_{\rm ad}$ for the emitting region
\citep[r$_{\rm ad}$ taken from][]{Church.vs.church04mn}.
If the emitting region is
0.1$\times$r$_{\rm ad}$ the densities increase by a factor of four. If
it is spread over the entire disk, the densities are a factor of
four less.
The electron densities that can be calculated from the
ionization parameters derived from the fits with photoionized absorbers for these systems in
\citet{ionabs:diaz06aa} yield a range of 8--0.2$\times$ 10$^{13}$
cm$^{-3}$ for the ionized absorbers in these systems. These values
are consistent with the maximum limits in the Comptonizing
regions. Then, the suggested warm absorbing regions may have similar
densities with the existing coronal regions.

The Compton radius in an accretion disk corona (ADC) with temperature
kT is r${\rm _C}$$\simeq$GMm${\rm _p}$/kT$_{plasma}$ \citep[cf.,][]{Church.vs.church04mn,b-church:aipc05}
where the hydrostatic
equilibrium fails. Outside this radius, the corona will dissipate
as a wind.  \citet{Church.vs.church04mn} shows that
the Compton radii for some LMXB dippers and the ADC radii calculated from the ingress and egress
timescales in their lightcurves are very similar.
We calculated the Compton radii for the four LMXBs
given the plasma temperatures from the fits with the {\tt CompTT} model
as follows :
\nineteen\ has r${\rm _C}$$\simeq$1.3$\times$$10^{10}$ cm, \fouru\
has r${\rm _C}$$\sim$1.8$\times$$10^{10}$ cm, \bigdip\
has r${\rm _C}$$\sim$3.9$\times$$10^{10}$ cm, and \thirteen\
has r${\rm _C}$$\sim$1.1$\times$$10^{9}$ cm.
The extent of the Comptonizing
regions, except for \thirteen, are large and
%in contrast with the calculated/expected values for
%ADCs in these sources \citep{Church.vs.church04mn} and
it could cover  13-77$\%$ of the disk (by area) given the statistical errors on the plasma temperature.
For  \thirteen\ the coverage of the ADC is only about 1$\%$ of the disk (by area), most likely a
small region confined to inner parts of the disk (given the size of calculated r${\rm _C}$).

We have also modeled the absorption in the broad-band INTEGRAL energy range with photoionized absorbers owing
to the detection of mainly iron absorption lines from these systems (other absorption lines have also been
detected). We were able to satisfactorily separate
dip and non-dip spectra from two of the systems and studied, for the first time, 4-200 keV spectra modeling
photoionized absorbers in these systems.
A comparison of the spectral results with the
XMM-Newton analysis in the 0.2-15.0 keV band (see Table~\ref{tab:properties})
yields \logxi\ values larger in the case of \nineteen\ and slightly smaller
in the case of \bigdip\ for both non-dip
spectra obtained from the time-averaged INTEGRAL data.
For dip spectra both sources show slightly higher ionization
parameter (compared with deepest dips in the XMM-Newton analysis).
This shows that INTEGRAL detects a more highly ionized
absorber in \nineteen\ (even higher than \bigdip) which was not recovered in the 0.2-15.0 keV analysis.
The neutral absorption of the highly ionized absorber is significantly lower
compared with the XMM-Newton results at all times.
%These discrepancies could be because the XMM-Newton observations
%were obtained at a time when the source was  
%in a particular low state in the case of \nineteen. 
The XMM-Newton analysis, also, assumes a photon index of $\Gamma=2$ for the ionizing
spectrum as assumed in this analysis,
but with an additional cut-off energy of 86 keV and 12 keV for \nineteen\ and \bigdip, respectively.
A comparison with the $Chandra$ (0.3-10.0 keV)
spectral results of the non-dip emission shows that INTEGRAL detects the same \logxi\
parameter.
A value of 3.73$\pm$0.13 is derived by \citet{1916:juett06apj} 
assuming a power law of -1 for the incident
ionizing flux.
%\citet{Iaria:apj06} also finds a \logxi\ of 4.15 . 
For \bigdip\ \citet{Iaria:aa07} measures
approximately a \logxi\ of 4.1 for the non-dip 
spectra which is higher then the INTREGRAL result.

\section{Summary and Conclusions}
\label{sect:Conclusions}

We have investigated 4--200 keV broad-band spectrum and temporal characteristics of
four dipping LMXB systems, \nineteen, \thirteen, \bigdip, and \fouru.
INTEGRAL has shown us for the first time
the time-averaged spectral parameters
of these systems together with the dip and non-dip
spectra.  This study also demonstrates for the first time,
a high sensitivity study of spectra and timing
characteristics of these four LMXBs above 20 keV
(simultaneously with 3-20 keV band data). INTEGRAL reveals the spectral characteristics
of the hottest regions in these dipping LMXBs together with the nature of the absorbers within the
systems. We find that the persistent emission spectral parameters of
\bigdip\ and \nineteen\ remain mostly the same
in and out the dipping intervals given the 4-200 keV energy range except for the changing absorption
below 10 keV for the two systems. \fouru\ shows a spectrum that is consistently the same, in and out the
X-ray dips between 4--200 keV given the INTEGRAL sensitivity.
INTEGRAL spectra are consistent with existence of persistent hot Comptonizing coronal regions
(ADCs), with plasma temperatures using Comptonized plasma models as
18.8$^{+10.9}_{-6.0}$ keV for \nineteen, 196.4$^{+28.5}_{-18.8}$ keV
for \thirteen, 9.2$^{+2.5}_{-1.9}$ keV for \bigdip\ and 4.9$^{+5.6}_{-1.9}$ keV for \fouru.
We caution that the ADC temperature for \thirteen\ is very high incomparison to
the rest and even within the LMXBs and the temperature limit of the {\tt CompTT} model
to satisfactorily represent the ADC falls readily after 50 keV \citep{farinelli:2008apj}.
Using our fits,
we derive a range of optical depth to Compton scattering $\tau$ in these regions as
 0.9$^{+0.7}_{-0.6}$  for \nineteen, 0.006$^{+0.001}_{-0.004}$ for \thirteen, 0.22$^{+4.21}_{-0.20}$
for \bigdip\ and 1.1$^{+2.3}_{-0.9}$ for \fouru. Given these results \thirteen\
has, by a large factor, the hottest ADC with the lowest optical depth.
None of the sources exhibit spectra consistent with a single model of emission except for  \thirteen
(i.e., a single power law model can be used to model the spectra).
Therefore, we stress that the spectra of these sources should never be fitted using a single model
of emisison regardless of the energy band (i.e., as in $Chandra$ or XMM-Newton bands). 
In addition, the four systems can not be fitted, alone, with a double
blackbody emission or double disk-blackbody emission models, either.
Fits to the INTEGRAL spectra  suggests a third component particularly of blackbody nature 
(due to the the detected different seed-in photon temparature compared with blackbody temperature)
that could be existent in the spectra.

For the first time, we detect $\sim$15$\%$, 40$\%$, and 45$\%$ modulation depth
(variation normalized to mean flux)
 in the 20-40 keV energy range
at the X-ray dip periods from \nineteen, \bigdip, and \fouru, respectively.
We belive this could be  as a result of scattering off of the
bulge at the accretion impact zone (fixed in the orbital plane).
Photoelectric absorption by material
at the impact zone is not as desirable an explanation since after 20 keV Comptonization and energy independent
scattering is expected to dominate. This is also supported by the large red noise (humps) found in the
average power spectra above 20 keV.

We have studied the effects of photionized absorbers in dipping LMXBs for the first time
in a  broad energy band 4-200 keV from \nineteen\ and \bigdip.
We find that the most highly ionized absorber is
in \nineteen\ with an ionization parameter \logxi\ of 3.8$^{+0.2}_{-0.4}$ although \bigdip\
has more luminosity. This
indicates the importance of system geometry and size. Moreover, inclusion of spectra from
a harder energy band compared with the XMM-Newton or Chandra data
help to constrain the absorber properties better.

The INTEGRAL spectrum reveals a small size accretion region
that is less than 1 $\%$ of the surface area of a typical neutron star 
in an LMXB derived from the blackbody fits for \nineteen\
and \thirteen. This is difficult to explain, alone, by the large amount of scattering in these systems.
On the contrary, accretion region sizes either about the size of a neutron star or 16-20 times larger (
than a neutron star in terms of accretion surface area)
are consistent with \bigdip\ and \fouru, respectively.
Such results from INTEGRAL
reveals different accretion geometries and emission regions  for these
dipping LMXB systems in contrast with the very similar modeling of such regions used
in previous studies (i.e., XMM-Newton, RXTE, Beppo-Sax).

The INTEGRAL broad-band hard X-ray and $\gamma$-ray
emission from the four dipping LMXB systems show that these
systems have different Comptonizing regions in size and in plasma temperature and has
different accretion geometries and emitting regions. We calculate that Comptonizing coronal regions
cover a larger fraction of the disk and are colder in comparison with
the conventional ADCs in three systems (i.e., \nineteen, \bigdip, \fouru), 
such regions vary with respect to the source and
are not only strongly
dependent on the central source luminosity. INTEGRAL detects that 
ADC in \thirteen\ is much smaller in size (much more hotter)
in comarison to the other three systems
and probably confined to the inner regions of the disk.  
% as was stressed earlier in  \citet{Church.vs.church04mn}.
We do not find a direct correlation of source luminosity with
plasma temperature in the ADCs of the four systems.
These extended highly ionized coronal regions (ADCs)
may act as an absorber to the incident radiation \citep{church:mnras05}.
As a result, we would be viewing these systems
through highly ionized absorbing regions at all times.
%The dipping
%activity is enhanced as a result of the absorption due to ionized or cold material
%at the  impact region of the accretion stream and/or the rim of the accretion disk which would yield 
%further absorption superimposed
%on the absorption from the coronal regions. 
On the other hand, INTEGRAL results  confirms the existence of warm absorbing
regions and that the absorbing material
%results confirm the XMM-Newton results for two of the dippers (\nineteen, \bigdipthe absorbing material 
looks less photoionized
during the main dips and more ionized as viewed in the persistent emission
intervals in a broad energy band of 4-200 keV for at least two of the dippers \nineteen\ and \bigdip.
These absorbing regions can be associated with an accretion disk atmosphere \citep[e.g., \exo:][]{Jimenez:2003apj}
However, the two systems are not well known X-ray line emitters.
It is unclear, at this stage, if
the absorber is the same in a given location, or if different absorbers are viewed during dipping and 
persistent emission episodes. It may be that it is a combined effect
of different absorbing regions  thoughout the disk (e.g., inner disk and outer disk)
superimposed on the spectra.

\begin{acknowledgements}

SB would like to thank Arvind Parmar, for introducing the topic of
dipping LMXBs and warm absorbers, and helpful discussions in constructing the paper.
I would also like to thank Mariano Mendez, Tim Oosterbroek, Maria Diaz Trigo, Ada Paizis, 
Lara Sidoli for discussions, help with data analysis and critical reading of
the manuscript.
This research was based on observations with INTEGRAL, an ESA project with
instruments and science data centre funded by ESA member states
(especially the PI countries: Denmark, France, Germany, Italy,
Switzerland, Spain), Czech Republic and Poland and with the
participation of Russia and the USA. SB acknowledges
an ESA fellowship. SB also acknowledges support from
from T\"UB\.ITAK, The Scientific and Technological Research Council
of Turkey,  through project 106T040.
%This research has made use of data obtained through the High Energy
%Astrophysics Science Archive Research Center Online Service, provided
%by the NASA/Goddard Space Flight Center.

\end{acknowledgements}

%------------------------------------
%       References
%------------------------------------

\bibliographystyle{apj}
\bibliography{mybib_up}

\end{document}